\documentclass[traditabstract]{aa} % for the abstract without structuration 
\usepackage{graphicx}
\usepackage{amsmath}
\usepackage{grffile}
\usepackage{siunitx}
\usepackage{txfonts}
\usepackage{graphicx}
\usepackage{caption}
\usepackage{natbib}
\usepackage{subfigure} %um Bilder nebeneinander zu plotten
\usepackage{hyperref}
\usepackage[utf8]{inputenc}
\usepackage{placeins}
\usepackage{scrextend}
\usepackage{mathtools, cuted}
\usepackage{rotating} %Um die Tabellen 90 Grad zu drehen verwende
\usepackage[normalem]{ulem}
\usepackage{multirow} %for \multirow command
\usepackage{booktabs} % to use \toprule, \midrule and \bootmrule in table

\newcommand{\Ms}{{\rm M}_\odot}

\newcommand{\mbh}{M_{\rm BH}}
\newcommand{\Mpc}{{\rm Mpc}}

\newcommand{\bd}{\begin{displaymath}}
\newcommand{\ed}{\end{displaymath}}
\newcommand{\be}{\begin{equation}}
\newcommand{\ee}{\end{equation}}
\newcommand{\beaa}{\begin{eqnarray*}}
\newcommand{\eeaa}{\end{eqnarray*}}
\newcommand{\bea}{\begin{eqnarray}}
\newcommand{\eea}{\end{eqnarray}}

\newcommand{\erg}{\mathrm{erg}}

\DeclareSIUnit\parsec{pc}
\DeclareSIUnit\lightyear{ly}
\DeclareSIUnit\year{yr}
\DeclareSIUnit\erg{erg}
\DeclareSIUnit\ster{ster}
\DeclareSIUnit\arcsec{arcsec}
\DeclareSIUnit\rad{rad}
\DeclareSIUnit\mag{mag}

\usepackage{color} %Text farbig machen
\definecolor{gruen}{cmyk}{0.35,0.01,0.80,0.1}

\definecolor{blue}{rgb}{0.0,0.0,1.0}
\definecolor{magenta}{rgb}{1.0,0.0,1.0}
\definecolor{orange}{rgb}{1.0,0.5,0.0}

\begin{document}

   \title{Strong lensing of tidal disruption events: Detection rates in imaging surveys}

  \titlerunning{Strongly lensed TDE detection rates}
   \author{K.~Szekerczes\inst{1}
                 \and    
          T.~Ryu\inst{1,2}
                    \and
          S.~H.~Suyu\inst{1,3,4}
                                \and
          S.~Huber\inst{1,3}
                    \and
            M.~Oguri\inst{5,6}
            \and
            L.~Dai\inst{7}
          }

   \institute{Max-Planck-Institut f\"ur Astrophysik, Karl-Schwarzschild Str. 1, 85741 Garching, Germany\\
              \email{kaitlyn@MPA-Garching.MPG.DE}
           \and 
           Physics and Astronomy Department, Johns Hopkins University, Baltimore, MD 21218, USA
         \and 
           Technical University of Munich, TUM School of Natural Sciences, Department of Physics, James-Franck-Stra\ss{}e~1, 85748 Garching, Germany
           \and
           Institute of Astronomy and Astrophysics, Academia Sinica, 11F of ASMAB, No.1, Section 4, Roosevelt Road, Taipei 10617, Taiwan
           \and
           Center for Frontier Science, Chiba University, 1-33 Yayoi-cho, Inage-ku, Chiba 263-8522, Japan;
           \and
           Department of Physics, Graduate School of Science, Chiba University, 1-33 Yayoi-Cho, Inage-Ku, Chiba 263-8522, Japan
           \and 
           Department of Physics, University of Hong Kong, Pokfulam Road, Hong Kong
}

%   \date{Received --; accepted --}

% \abstract{}{}{}{}{} 
% 5 {} token are mandatory
 
  \abstract
{Tidal disruption events (TDEs) are multi-messenger transients in which a star is tidally destroyed by a supermassive black hole at the center of galaxies. The Rubin Observatory Legacy Survey of Space and Time (LSST) is anticipated to annually detect hundreds to thousands of TDEs, such that the first gravitationally lensed TDE may be observed in the coming years. Using Monte-Carlo simulations, we quantify the rate of both unlensed and lensed TDEs as a function of limiting magnitudes in four different optical bands ($u$, $g$, $r$, and $i$) for a range of TDE temperatures that match observations. Dependent on the temperature and luminosity model, we find that $g$ and $r$ bands are the most promising bands with unlensed TDE detections that can be as high as ${\sim}10^{4}$ annually. By populating a cosmic volume with realistic distributions of TDEs and galaxies that can act as gravitational lenses, we estimate that a few lensed TDEs (depending on the TDE luminosity model) can be detected annually in $g$ or $r$ bands in the LSST survey, with TDE redshifts in the range of ${\sim}0.5$ to ${\sim}2$. The ratio of lensed to unlensed detections indicates that we may detect ${\sim}1$ lensed event for every $10^{4}$ unlensed events, which is independent of the luminosity model. The number of lensed TDEs decreases as a function of the image separations and time delays, and most of the lensed TDE systems are expected to have image separations below ${\sim}3 \arcsec$ and time delays within ${\sim}30$ days.
At fainter limiting magnitudes, the $i$ band becomes notably more successful. These results suggest that strongly lensed TDEs are likely to be observed within the coming years and such detections will enable us to study the demographics of black holes at higher redshifts through the lensing magnifications.}

   {}
   {}
   {}
   {}


   \maketitle
%
%________________________________________________________________

\section{Introduction}

Tidal disruption events (TDEs) occur when a star passes sufficiently close to
a supermassive black hole (SMBH) such that the tidal forces of the black hole (BH) overcome the star's self gravity, and the star is subsequently disrupted \citep{Hills1988,Rees1988}, generating a flare that is detectable on the time scale of a few months to a few years \citep[e.g.,][]{Gezari2021}. Since quiescent BHs can create TDEs, TDEs can serve as a promising tool for probing these dormant BHs. The population studies of TDEs in dwarf galaxies will provide a better understanding of the BH mass function. Currently, on the order of 100 TDE candidates have been detected with a typical redshift of $z\lesssim 0.2$, but this number will exponentially grow to possibly more than thousands with detections by ongoing (e.g., Zwicky Transient Facility \citep[ZTF;][]{ZTF} and eROSITA \citep{eROSITA}) and upcoming surveys (e.g., Rubin Observatory Legacy Survey of Space and Time \citep[LSST;][]{LSST} and Ultraviolet Transient Astronomy Satellite \citep[ULTRASAT;][]{ULTRASAT}). 

Although the detections of TDEs so far have been confined within the nearby Universe, TDEs at comological distances could be detected in the future with observing instruments with deeper fields. For that case, it is possible these events can be gravitationally lensed by galaxies at lower redshifts, magnifying the brightness of the TDEs. Such lensing effects would enable the detections of TDEs at even higher redshifts, which would allow us to study the TDE rates and the BH mass function at the lower mass end and at earlier cosmic time. Additionally, through studying the microlensing of TDEs, we can obtain an independent constraint on the size of the TDE emitting region which can help to identify its emission mechanism \citep[as demonstrated in studies of active galactic nuclei (AGNs) with their accretion disk sizes measured through microlensing, e.g., ][]{Kochanek_2004, Schmidt_2010, Blackburne_2014}. 

In fact, the strongly lensed detection rates have been estimated for quasars (a type of AGN) and supernovae for various surveys (e.g., LSST and Supernova Legacy Survey (SNLS)) by \citet{Oguri_2010}. Specifically, \citet{Oguri_2010} populate a region of their simulation with lens galaxies and either quasars or supernovae by randomly assigning redshifts to both objects and a magnitude to the source based on their luminosity and mass functions. They then compute the lensing effects for every source in the simulation and determine if the strongly lensed system will be detected by the particular survey. Through these simulations, \citet{Oguri_2010} estimated ${\sim}3000$ lensed quasars and $\sim$100 lensed supernova will be detectable with LSST. TDEs, although rarer than supernovae and quasars, could potentially be strongly lensed and detected in LSST as well.

As the number of detected TDEs is expected to grow in the coming years, it is important to quantify how many lensed TDEs may be detected by full sky surveys.
We focus on calculating the strongly lensed TDE detection rates in this paper.
As a first step, using the code developed by \citet{Oguri_2010}, we estimate the strongly lensed TDE detection rates over a wide range of limiting magnitudes and blackbody temperatures assuming two luminosity models which bracket the range of observed TDE luminosities. We then choose the blackbody temperature that best corresponds with current ZTF observations. Then, using the chosen temperature model, we compute the strongly lensed TDE detection rates for both luminosity cases. We use the range of detection rates from the two luminosity models to provide a bound on the strongly lensed detection rate.

Recently, \citet{Chen+2024} have independently estimated the rates of strongly lensed TDEs through simulations of TDE light curves and computations of lensing probabilities.  Our results on the lensed TDE rates are overall consistent with the lower end of the range predicted by \citet{Chen+2024}, although the details of models and criteria used for detection are different, which we discuss further in Section \ref{sec: discussion}.

This paper is organized as follows. We provide the descriptions for our methodology and present the estimated unlensed TDE rates in Section \ref{sec: Rates of unlensed TDE}. Then, we explain our strategies for estimating lensed rates in Section \ref{sec: Rates of lensed TDE}; we give the estimates of the lensed TDE rates, and we show the distributions of the lens properties from the resulting strongly lensed systems. We discuss some of the caveats of this work in Section \ref{sec: discussion}, and we conclude in Section \ref{sec: summary}.

\section{Detection rates of unlensed TDEs}
\label{sec: Rates of unlensed TDE}
In this section, we compute the rates of TDEs based on theoretical models that are matched to observations. We begin in Section \ref{sec:unlensedTDEs:overview} with an overview of our computation of the rates, and detail the ingredients for this computation in Sections \ref{subsec:luminosity} to \ref{subsec:conversion2}. Our results of the detection rates of TDEs are presented in Section \ref{subsec:unlensedTDEs:results}.

\subsection{Overview of method}
\label{sec:unlensedTDEs:overview}
We calculate the detection rates of unlensed TDEs for each $X$ band ($X$ = $u$, $g$, $r$, $i$) by performing the following integral, 

\begin{equation}
    N_{\rm TDE,X} = \int_{0}^{z_{\rm max}} \int_{m_{\rm max}}^{m_{\rm min}}\psi_{\rm TDE}(m_{\rm X}',z)\,\mathrm{d}m_{\rm X}'\, \frac{{\rm d}V(z)}{{\rm d}z}\, {\rm d}z
    \label{eq:unlensed_rate},
\end{equation}
where $\psi_{\rm TDE}$ is the TDE luminosity function as a function of magnitude $m_\mathrm{X}$ detected in the $X$ band and $\mathrm{d}V/\mathrm{d}z$ is the cosmological volume element (see Equation 8 in \citet{Oguri_2010}) assuming ($\Omega_{m}$, $\Omega_{\Lambda}$) = (0.3, 0.7) and $H_{0}=72 \, {\rm km}~{\rm s}^{-1}{\rm Mpc}^{-1}$. 
The upper bound $m_\mathrm{min}$ is determined by the brightest event, and $m_\mathrm{max}$ is set by the smaller of either the faintest event or the filter magnitude limit. The $z_{\rm max}$ bound is determined by the TDE with the highest luminosity contained within the filter magnitude limit. To compute $\psi_{\rm TDE}(m_{\rm X})\mathrm{d}m_{\rm X}$, we first construct $\psi_{\rm TDE}(M_{\rm BH})$ as a function of BH mass, $\mbh$, from the BH mass function, $\phi_{\rm BH}$, and the TDE rate per galaxy, $\Gamma$ (see Section \ref{subsec:luminosity}). Then, we convert $\psi_{\rm TDE}(M_{\rm BH})$ into $\psi_{\rm TDE}(L)$ as a function of luminosity, $L$, assuming two different TDE luminosities  (see Section \ref{subsec:conversion1}). Finally, we convert  $\psi_{\rm TDE}(L)$ into $\psi_{\rm TDE}(m_{\rm X})$ for different bands assuming black body emissions (see Section \ref{subsec:conversion2}).

We compute these unlensed detection rates for a range of constant temperatures and compare with current ZTF detections to determine the model that best corresponds with observations. Additionally, we show how the rates evolve with a variable magnitude cutoff assuming a particular survey area of 20,000 $\mathrm{deg}^{2}$. In this way, the annual detection rates in any imaging survey can be estimated given its magnitude limit.

\subsection{Luminosity function - $\psi_{\rm TDE}\left(\mbh \right){\rm d}\mbh$}\label{subsec:luminosity}
\begin{figure}[h!]
    \centering
    \includegraphics[scale =0.58, trim={0.1in 0.1in 0.08in 0.1in}, clip]{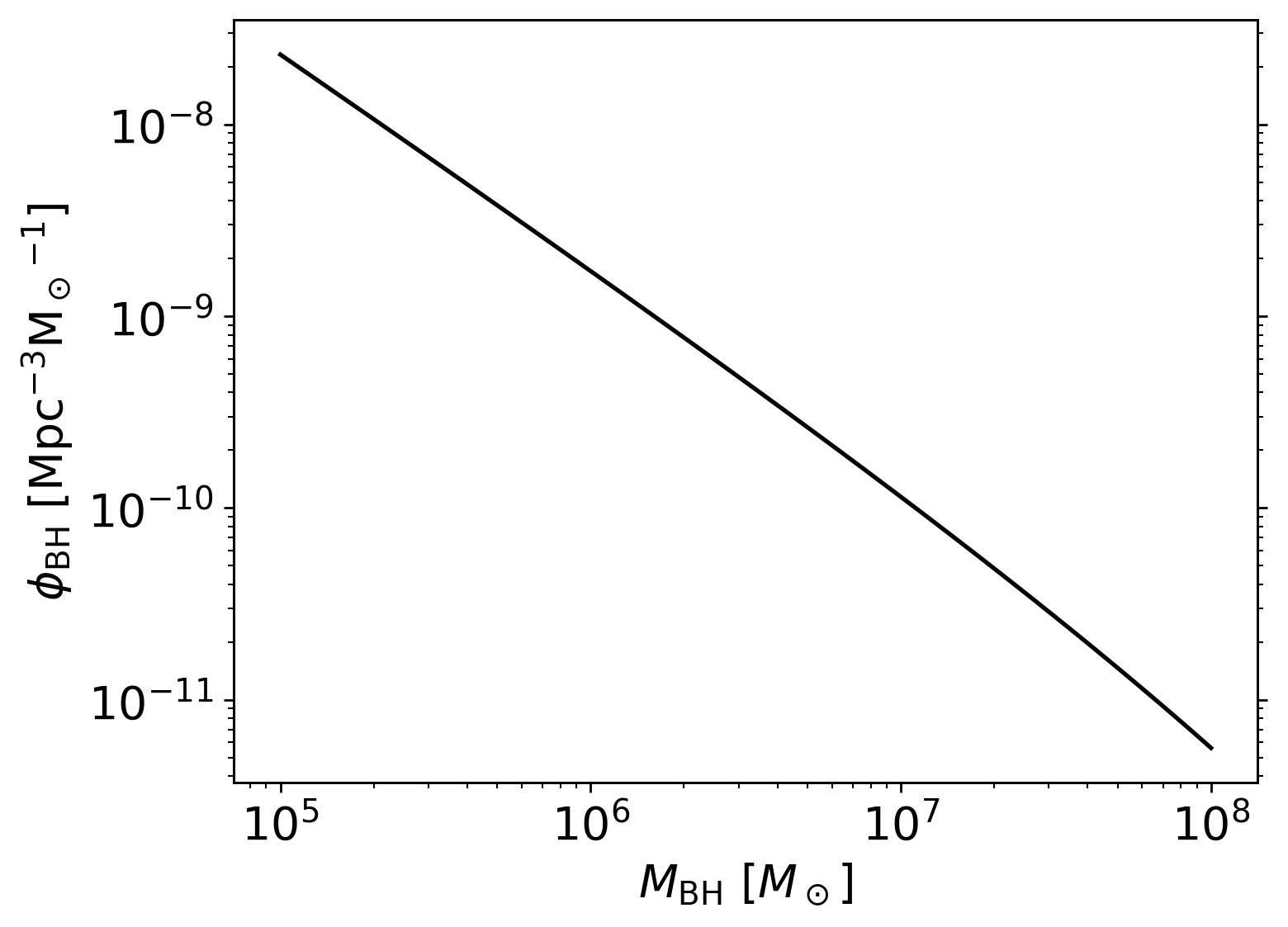}
    \caption{BH mass function $\phi_{\rm BH}$ for the range of $10^{5}$ to $10^{8}\,\Ms$.}
    \label{fig:bhmf}
\end{figure}

We first express the luminosity function, $\psi_{\rm TDE}$, as a function of BH mass, $\mbh$, by multiplying the BH mass function, $\phi_{\rm BH}$, and the TDE occurrence rate, $\Gamma$.
We employ a local BH mass function derived from early-type galaxies within 30 Mpc \citep{Gallo_2019, Wong_2022}, 

\begin{align}\label{eq:bhmf}
    \log_{10}\left(\frac{\phi_{\rm BH}}{\Mpc^{3}\,\Ms}\right) & = -9.82 -1.10\log_{10}\left(\frac{\mbh}{10^{7}\,\Ms}\right)\nonumber\\
     &- \left[\frac{\mbh}{128 \times 10^{7}\,\Ms}\right]^{\left(\frac{1}{\ln(10)}\right)},
\end{align}
which gives the number of BHs per volume for a given $\mbh$ range. We assume that $\phi_\mathrm{BH}$ does not evolve with redshift. This BH mass function is shown in Figure \ref{fig:bhmf}. One can derive a redshift dependent BH mass function by combining a galaxy mass function and a galaxy-BH scaling relation. However, galaxy scaling relations may break down near the low-mass end, and the occupation fraction of BHs in high-$z$ dwarf galaxies is uncertain. Since low-mass SMBHs would create the majority of TDEs, we employed the BH mass function that takes into account the occupation fraction although it is valid only for the local universe. We will explore the dependence of the lensed TDE rates on the redshift evolution of the BH mass function in our future work.

We take the TDE occurrence rate, $\Gamma$, from \citet{Pfister_2020},

\begin{equation}\label{eq:rate}
     \Gamma(\mbh) = 10^{-4.5}\mathrm{yr}^{-1} \mathrm{galaxy}^{-1} \left(\frac{M_{\rm BH}}{10^{6}\,\Ms}\right)^{-0.14},
\end{equation}
which yields for a given BH mass the number of TDEs per galaxy per year. This rate is in agreement with current observed TDE rates. This TDE rate per galaxy is also valid within the local universe and assumed to not evolve with redshift. Then, assuming each galaxy harbors one massive BH at its center, we write $\psi_{\rm TDE}(\mbh)$ as, 

\begin{equation}\label{eq:tderate}
    \psi_{\rm TDE}\left(\mbh\right) =\phi_{\rm BH}\Gamma= 10^{-13.22}  \left(\frac{M_{\rm BH}}{10^{6}\,\Ms}\right)^{-1.24}10^{ -\left[\frac{M_{\rm BH}}{128\times 10^{7}\,\Ms}\right]^{\left(\frac{1}{\ln(10)}\right)}},
\end{equation}
From Equation (\ref{eq:tderate}), $\psi_{\rm TDE}\,{\rm d}\mbh$ gives the number of TDEs per year per volume within a BH mass range of $M_{\rm BH}$ to $M_{\rm BH} + {\rm d}M_{\rm BH}$. We consider the range of $\mbh$ within $10^{5}\,\Ms$ and $10^{8}\,\Ms$ based on current observational constraints such as, e.g., the TDEs from ZTF \citep{Yao+2023}.  Above the upper limit of ${\sim}10^{8}\,\Ms$ in the BH mass, stars are swallowed whole by the central BH rather than being tidally disrupted.

\subsection{Conversion from 
$\psi_{\rm TDE}(\mbh)$ to $\psi_{\rm TDE}(L)$}
\label{subsec:conversion1}

\begin{figure}[h!]
    \centering
    \includegraphics[scale =0.6, trim={0.1in 0.1in 0.08in 0.1in}, clip]{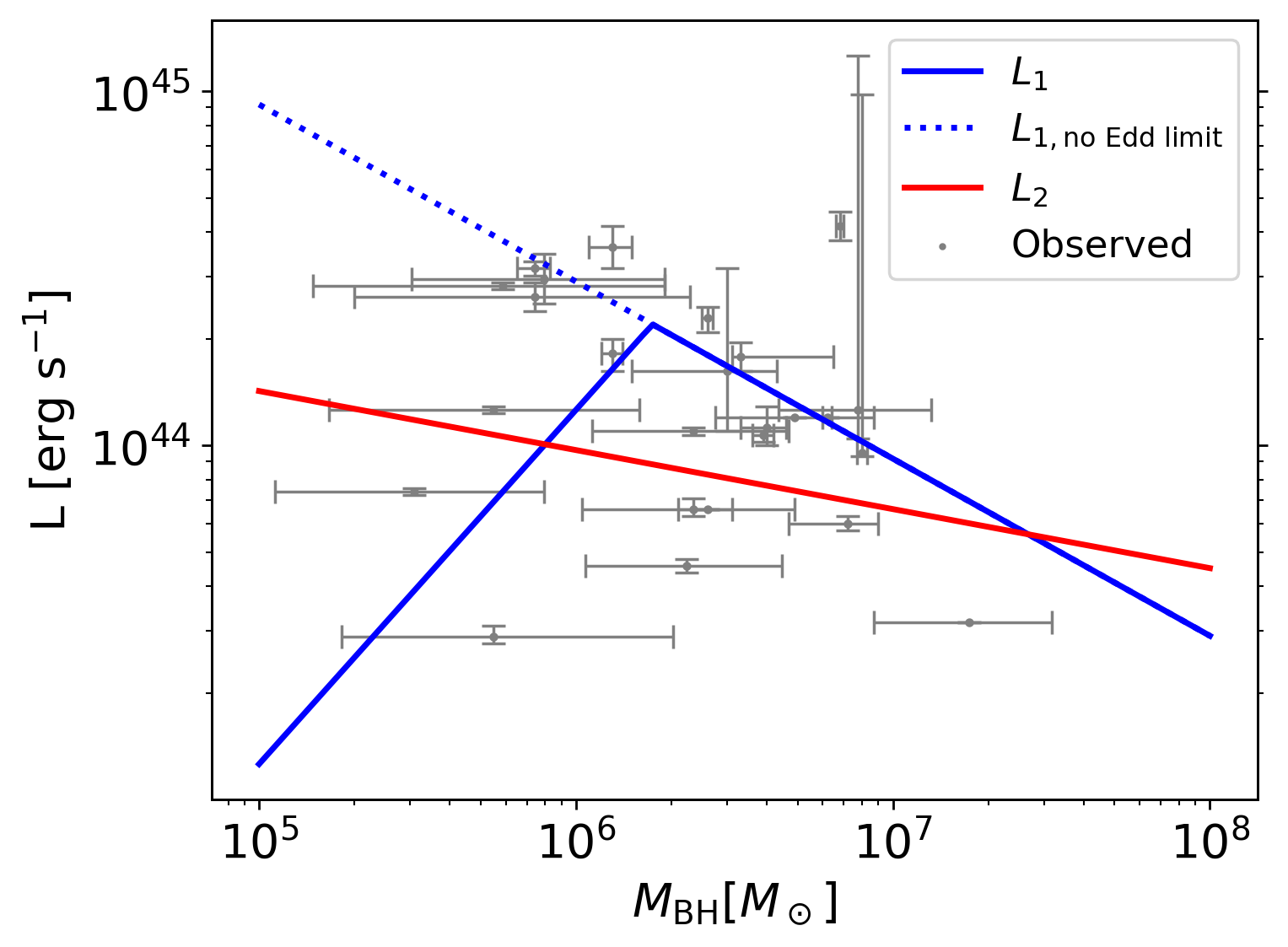}
    \caption{Two luminosities of TDEs, $L_{1}$ \citep[Equation (\ref{eq:L1_simplified}),][]{Thomsen_2022} and $L_{2}$ \citep[Equation (\ref{eq:L2}),][]{2020ApJ...904...73R} as a function of $M_{\rm BH}$ overlaid with observed optical TDEs using grey dots and error bars (see Table A.1 in \citep{Wong_2022} and Table 1 in \citep{2020ApJ...904...73R} and references therein). The dotted blue line conveys $L_{1}$ as given by Equation (\ref{eq:L1_simplified}), whereas the solid blue line show the Eddington limited $L_{1}$.}
\label{fig:lum}
\end{figure}

We now find the expression for $\psi_{\rm TDE}(L)$ from the relation  $\psi_{\rm TDE}(L)=\psi_{\rm TDE}(M_{\rm BH}(L))\,{\rm d}M_{\rm BH}/{\rm d}L$. To do this, we consider two different TDE luminosity expressions, which allows us to assess how the differing assumptions in turn alter the detection rate. The first, which we denote as $L_{1}$, is an observationally driven upper limit that about 1\% of the rest mass energy of the stellar fallback material produced in a disruption of $0.1\,\Ms$ main-sequence star turns into radiation with an efficiency of $\eta = 0.01$ \citep{Thomsen_2022} ,
\begin{align}\label{eq:L1}
    L_{1}(\mbh) &= \eta \dot{M}_{\rm fb} c^{2},
\end{align}
where $\dot{M}_{\rm fb}$ is the mass fallback rate of the most bound debris, and $c$ is the speed of light\footnote{The stellar mass of 0.1$\,\Ms$ and efficiency of $\eta=0.01$ are values that yield a luminosity-mass relation which is compatible with the more luminous TDEs (Figure \ref{fig:lum}).}.
Using the expression of $\dot{M}_{\rm fb}$ \citep[see Equation (16) in ][]{2020ApJ...904...98R}, we obtain,
\begin{align}\label{eq:L1_simplified}
    L_{1}(\mbh) &= 2.9\times 10^{44} \mathrm{erg\,s^{-1}}\left(\frac{\eta}{0.01}\right)\left(\frac{\mbh}{10^{6}\,\Ms}\right)^{-1/2}.
\end{align}
The second luminosity, denoted $L_{2}$, is driven by self-crossing shocks between debris streams \citep{2020ApJ...904...73R}, 
\begin{align}\label{eq:L2}
    L_{2}(\mbh) &= \frac{G \mbh\dot{M}_{\rm fb}}{a},\nonumber\\
       &= 9.7\times10^{43} \mathrm{erg\,s^{-1}} \left(\frac{\mbh}{10^{6}\,\Ms}\right)^{-1/6},
\end{align}
where the apocenter distance of the most bound debris $a$ is estimated assuming a $1\,\Ms$ star. \citep{2020ApJ...904...98R} corrected the luminosity using a correction factor that incorporates relativistic effects and realistic stellar internal structure. In this work, we only take the correction for the internal structure and neglect the term for relativistic effects. It is because this is only relevant at higher $M_{\rm BH}$, which constitutes an insignificant fraction of the detected population. 
In Figure \ref{fig:lum}, the luminosity models are shown for the BH mass range that we consider, and these are plotted with optically observed TDEs to show how they correspond with observed data. As we later mention in Section \ref{subsec:unlensedTDEs:results}, we limit the luminosity for $L_{1}$ so as to produce more realistic rates, so we show the Eddington limited $L_{1}$ (solid blue) in this Figure. This Eddington limited $L_{1}$ model is used throughout the rest of this work; the dotted blue line for the $L_{1}$ without an Eddington limit is shown for reference and not used for computing detection rates.

Using Equations (\ref{eq:L1_simplified}) and (\ref{eq:L2}), we compute ${\rm d}\mbh/{\rm d}L$ for both $L_{1}$ and $L_{2}$,
\begin{align}\label{eq:dmdl1}
     \frac{\mathrm{d}\mbh}{\mathrm{d}L_{1}} &\simeq -3.35\times10^{-4} {\rm erg}^{-1}~{\rm s} ~{\rm g} \left(\frac{L_{1}}{10^{44}~{\rm erg}~{\rm s}^{-1}}\right)^{-3},\\\label{eq:dmdl2}
     \frac{\mathrm{d}\mbh}{\mathrm{d}L_{2}} &\simeq -10^{-4} {\rm erg}^{-1}~{\rm s} ~{\rm g} \left(\frac{L_{2}}{10^{44}~{\rm erg}~{\rm s}^{-1}}\right)^{-7}.
\end{align}
So we can find an expression for $\psi_{\rm TDE}(L)$ using Equations (\ref{eq:L1_simplified}) and (\ref{eq:dmdl1}) for $L_{1}$ and  Equations (\ref{eq:L2}) and (\ref{eq:dmdl2}) for $L_{2}$. Note that ${\rm d}M_{\rm BH}/{\rm d}L$ has a negative sign because $L$ is anti-correlated with $\mbh$ (see Equations (\ref{eq:L1_simplified}) and (\ref{eq:L2})). This also means that the negative sign of ${\rm d}M_{\rm BH}/{\rm d}L$ will be cancelled by integrating over $L$ from a larger $L$ to a smaller $L$, corresponding to integrating from the lower bound of $\mbh$ to the upper bound, giving positive $\Psi_{\rm unlensed}$. 

\subsection{Conversion from $\psi_{\rm TDE}(L)$ to $\psi_{\rm TDE}(m_{\rm X},z)$}\label{subsec:conversion2}

We now need to convert $\psi_{\rm TDE}(L)$ into $\psi_{\rm TDE}(m_{\rm X},z)$  using the relation $\psi_{\rm TDE}(m_{\rm X},z) = \psi_{\rm TDE}(L(m_{\rm X},z))\, {\rm d}L/{\rm d}m_{\rm X}$. 
To do this, we need to find an expression of the magnitude $m_{\rm X}$ as a function of $L$. We assume a blackbody spectrum with a constant temperature $T$, giving the spectral intensities, in units of power per solid angle per area per wavelength, 
in the source frame as, 
\begin{equation}\label{eq:bb}
    I(T,\lambda_{\rm e}) = \frac{2hc^{2}}{\lambda_{\rm e}^{5}} \frac{1}{e^{\frac{hc}{\lambda_{\rm e} kT}}-1},
\end{equation}
where $h$ is the Planck constant, $k$ the Boltzmann constant, and $\lambda_{\rm e}$ the emitted wavelength. It follows that the observed flux is, 
\begin{equation}
    F_{\lambda , o} = \frac{I(T, \lambda_{\rm o})L}{4 D_{\rm L}^{2}(1+z)\sigma T^{4}},
\end{equation}
where $D_{\rm L}$ is the luminosity distance of the source, $\lambda_{\rm o}$ the observed wavelength, $\lambda_{\rm o}=(1+z)\lambda_{\rm e}$, $z$ the source redshift, and $\sigma$ the Stefan-Boltzmann constant.

\begin{figure}[h!]
    \centering
\includegraphics[width =8.cm]{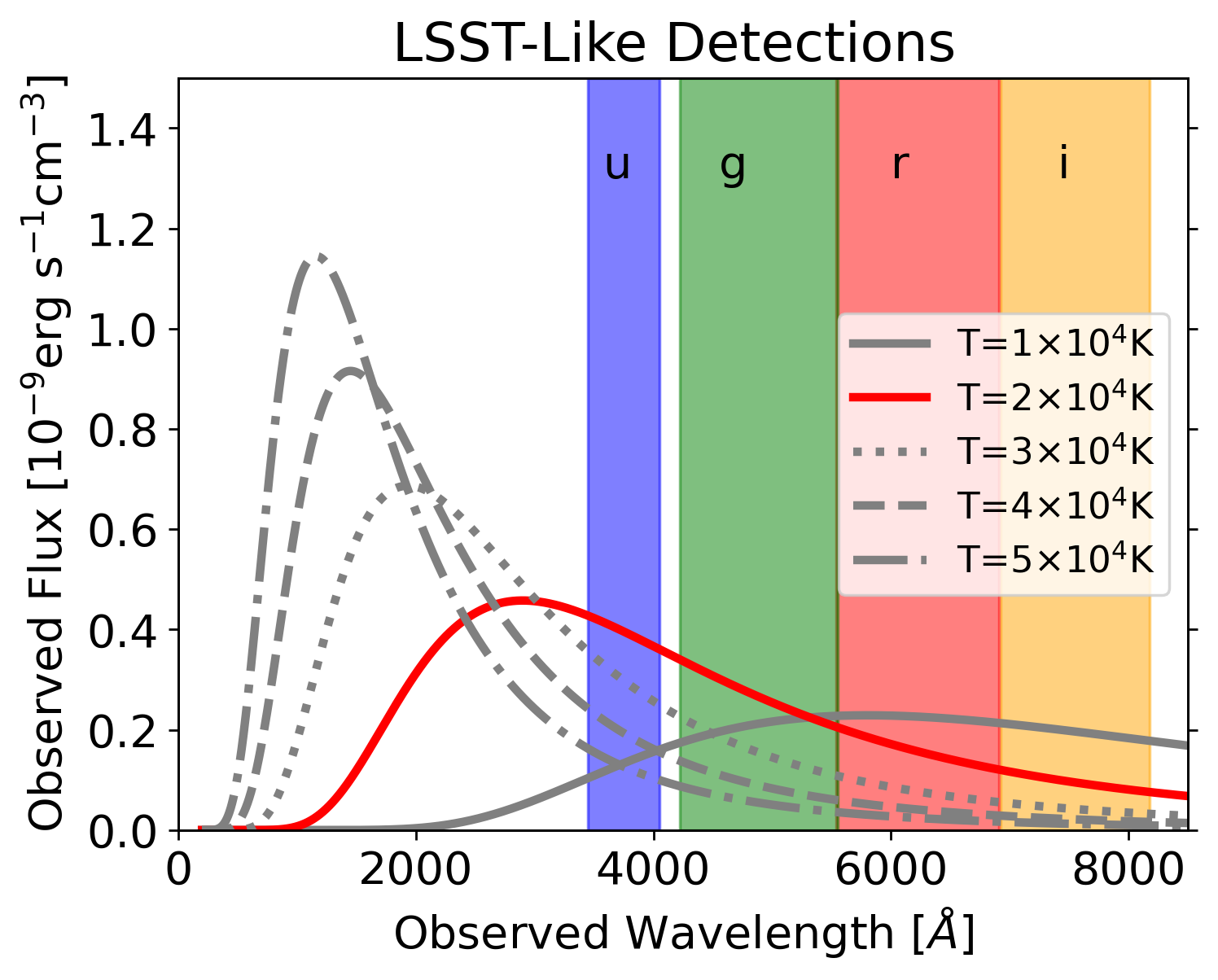}
\includegraphics[width =8.cm]{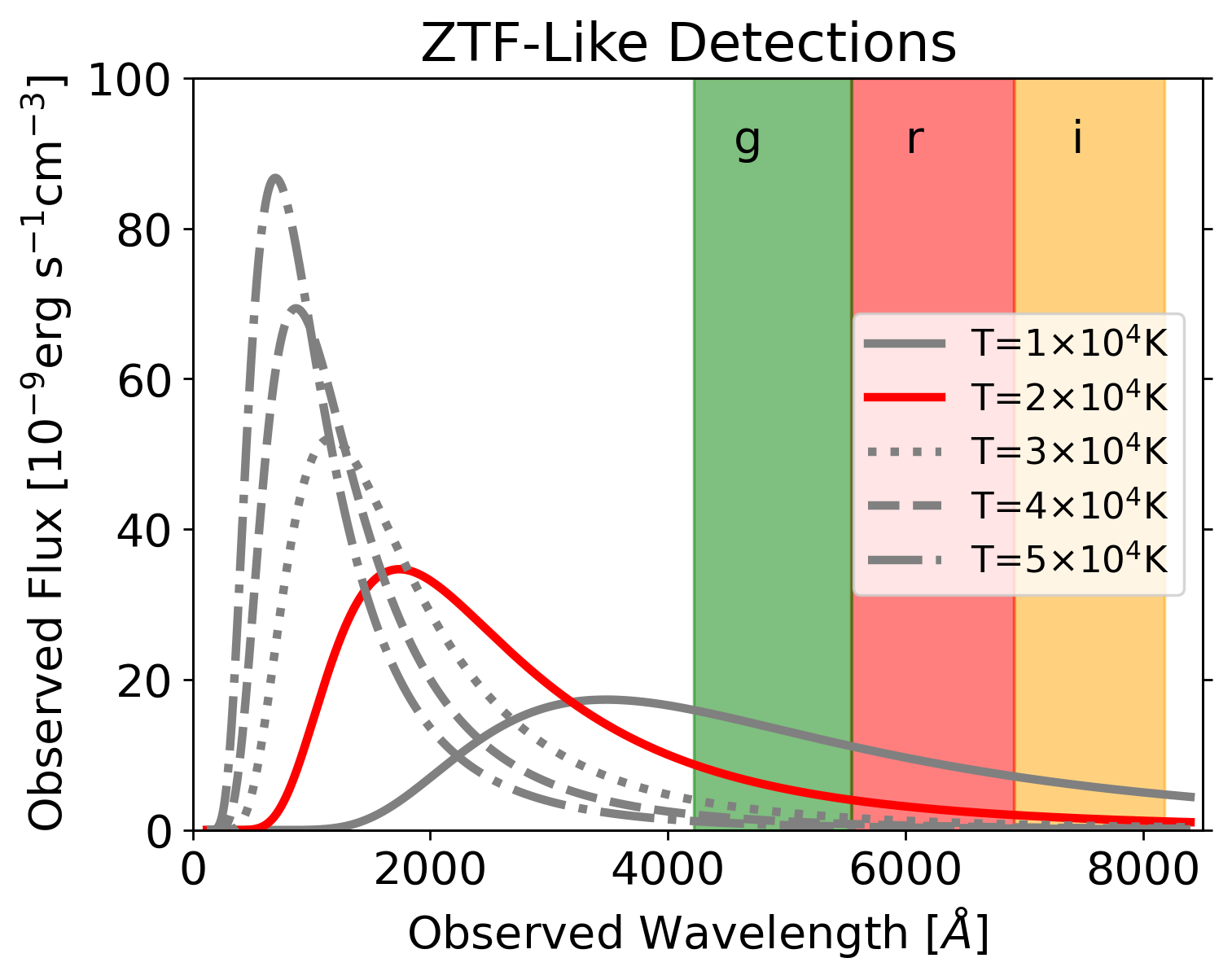}
\caption{Observed flux for a TDE with $L = 10^{44}\, \mathrm{erg}~\mathrm{s^{-1}}$ at $T = 1$, $2$, $3$, $4$ and $5 \times 10^{4}$ K assuming typical redshifts given our unlensed computations for LSST and ZTF which are $z = 1.0$ and $z = 0.2$, respectively.}
\label{fig:l2_flux}
\end{figure}

As driven by observations, we assume that the TDEs have a relatively constant temperature with a temperature ranging from $(1 - 5) \times 10^{4}$ K \citep[e.g.,][]{2021ARA&A..59...21G}. Figure \ref{fig:l2_flux} shows the observed flux for five different temperatures within our considered temperature range. A typical luminosity of $L = 10^{44} ~\mathrm{erg}~\mathrm{s^{-1}}$ is chosen to convey these flux trends for multiple temperatures, and the redshifts correspond to median source redshifts for detectable unlensed TDEs by LSST and ZTF which we will show in Section \ref{subsec:unlensedTDEs:results}. We indicate the wavelength range of each band for both surveys using different colors. The observed wavelength at the peak of the flux $\lambda_\mathrm{o,peak}$ follows the Wien's displacement law, $\lambda_\mathrm{o,peak}$ $\propto (1+z)/T$. For LSST, the flux peaks fall at an observed wavelength of $\lambda_\mathrm{o,peak} = 5799$, $2899$, $1933$, $1449$ and $1160\,\AA$
for $T = 1$, $2$, $3$, $4$ and $5 \times 10^{4}$ K, respectively. For ZTF, the fluxes peak at $\lambda_\mathrm{o,peak} = 3480$, $1740$, $1160$, $870$ and $696\,\AA$ taken over the same temperature range as LSST.

We can use the observed flux to calculate the apparent magnitude $m_{\rm X}$ for a particular filter $X$ ($X = u$, $g$, $r$, and $i$ bands) within the filter wavelength limits $\lambda_{\rm min}$ to $\lambda_{\rm max}$ \citep{2012PASP..124..140B}.

\begin{align}
    m_{\rm X}(L) &= -2.5 \log_{10}\left(\frac{ \int_{\lambda_{\rm min}}^{\lambda_{\rm max}} \lambda_{\rm o}^{'} S_{X}(\lambda_{\rm o}^{'})F_{\lambda_{\rm o}^{'} , o} d\lambda_{\rm o}^{'}}{ \int_{\lambda_{\rm min}}^{\lambda_{\rm max}} S_{X}(\lambda_{\rm o}^{'}) \frac{c}{\lambda_{\rm o}^{'}} d\lambda_{\rm o}^{'}} \right) -48.6,\nonumber\\
                 & = -2.5\log_{10}(\Xi(T, \nu_{\rm max}, \nu_{\rm min},z) L) - 48.6,
    \label{eq:mag}
\end{align}
where  $S_{\rm X}(\lambda)$ is the transmission function for filter $X$, and $\Xi(T, \nu_{\rm max}, \nu_{\rm min},z)$ is the collection of all terms from the integration except $L$ so as to simply show the relation between magnitude and luminosity.  We provide the apparent magnitude in terms of frequency as performing the integration over frequency allows us to obtain an analytic expression for the apparent magnitude. As shown in Figure 1 of \citet{Huber_2021}, the transmission functions are well approximated by top-hat functions for the different wavelength ranges. We assume a top-hat function for these filter functions with the wavelength ranges approximated at $\lambda_\mathrm{u} = 3450 - 4050 ~\AA$, $\lambda_\mathrm{g} = 4220 - 5550 ~\AA$, $\lambda_\mathrm{r} = 5540 - 6910 ~\AA$ and $\lambda_\mathrm{i} = 6920 - 8180 ~\AA$. These approximations are determined by fitting the LSST filter functions, but these are also reasonable approximations for the ZTF filters \citep[see Figure 2 in ][]{Bellm_2018}. Now, $\Xi$ can be expressed as 

\begin{align}\label{eq:xi}
&\Xi(T, \nu_{\rm max}, \nu_{\rm min},z) =\left[\frac{(1 + z )}{4D_{L}^{2}\sigma T^{4}c^{2}h^{2}} \right]\left[\ln\left(\nu \right) \Big|_{\nu_{\rm min}}^{\nu_{\rm max}}\right]^{-1}\nonumber \\
   &\times \Big[
    2hkT \nu (1+z)
    \left(h \nu (1+z)
    \ln\left[1 - e^{\frac{-h \nu (1+z)}{kT}} \right]- 2kT {\rm Li}_{2}\left(e^{\frac{-h \nu(1+z)}{kT}}\right)\right)\nonumber\\
    &- 4(kT)^{3}{\rm Li}_{3}\left(e^{\frac{-h \nu (1+z)}{kT}}\right)\Big|_{\nu_{\rm min}}^{\nu_{\rm max}}\Big],
\end{align}
where ${\rm Li_{s}}(z)$ is a polylogarithm of order $s$ and argument $z$. 

\begin{figure}[h!]
    \centering
\includegraphics[width = 8.cm]{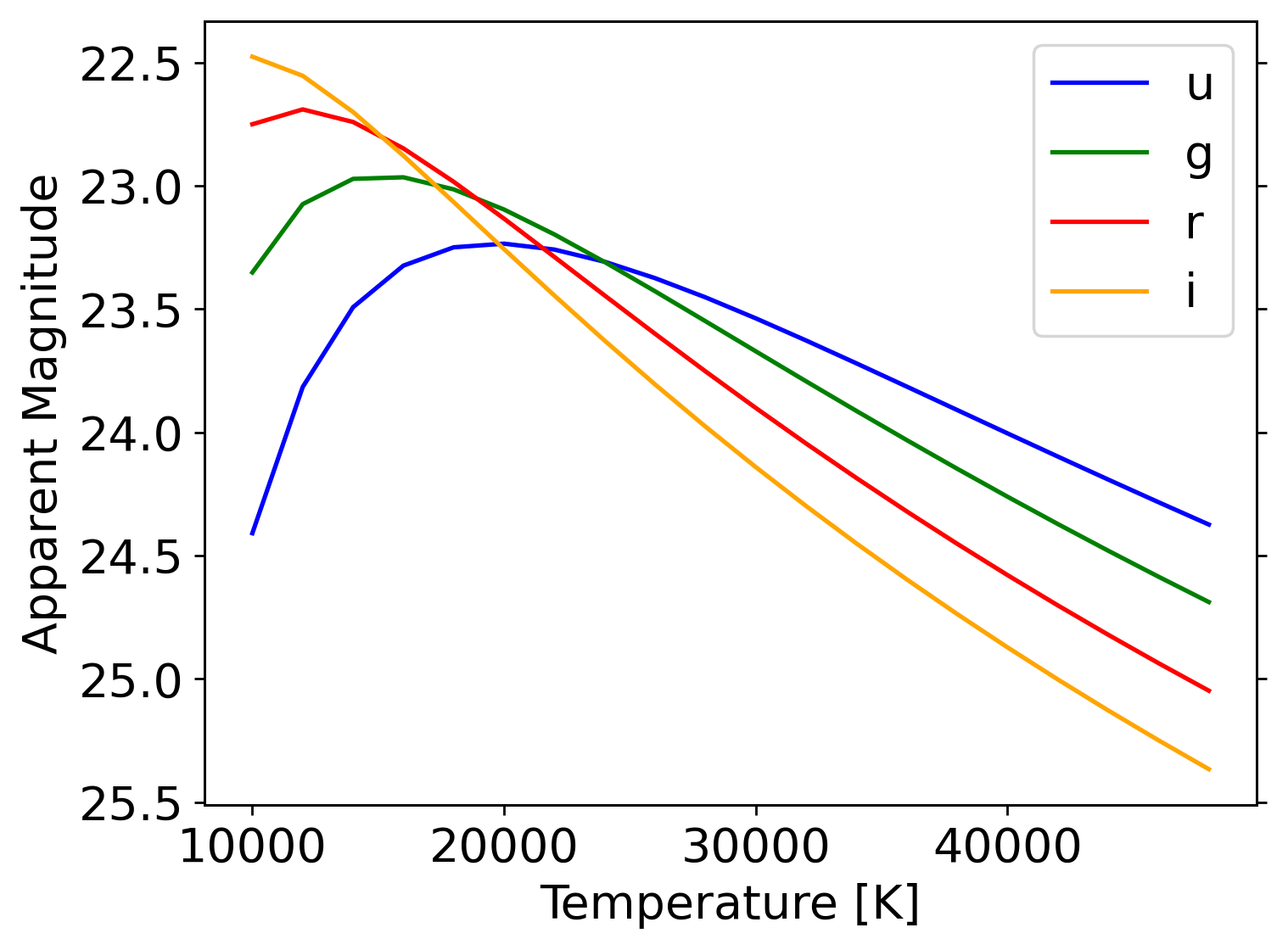}
\caption{Apparent magnitudes for $u$, $g$, $r$, and $i$ bands as a function of temperature at a redshift of $z=1.0$ and assuming $L= 10^{44}\,\rm{erg}~\rm{s^{-1}}$.}
\label{fig:l2_mag}
\end{figure}

Figure \ref{fig:l2_mag} shows how the apparent magnitude changes with temperature, at $z = 1.0$. For $i$ band, as temperature increases, the apparent magnitude becomes progressively fainter. Whereas the trend is initially the opposite for the other three bands, the apparent magnitude becomes brighter before dimming, yet the $u$ and $g$ bands still remain brighter than the other two bands for $T > 2.2 \times 10^{4}$ K. 
As apparent magnitude directly depends on the observed flux, these trends are a result of the peak of the observed flux moving out of the $i$ and $r$ bands and into the bands corresponding to shorter wavelengths, $u$ and $g$, for an increasing temperature as seen in Figure \ref{fig:l2_flux}. Given these magnitude trends for LSST-like limiting magnitudes, we can expect that at the lowest temperature, the $i$ and $r$ bands will have the highest detections. In contrast, the mid-range to higher temperatures will result in $g$ band having the highest detections; $u$ band, while bright at these temperatures as well, is less sensitive which will inhibit its detections.

\begin{figure}[h!]
    \centering
\includegraphics[width = 8.cm]{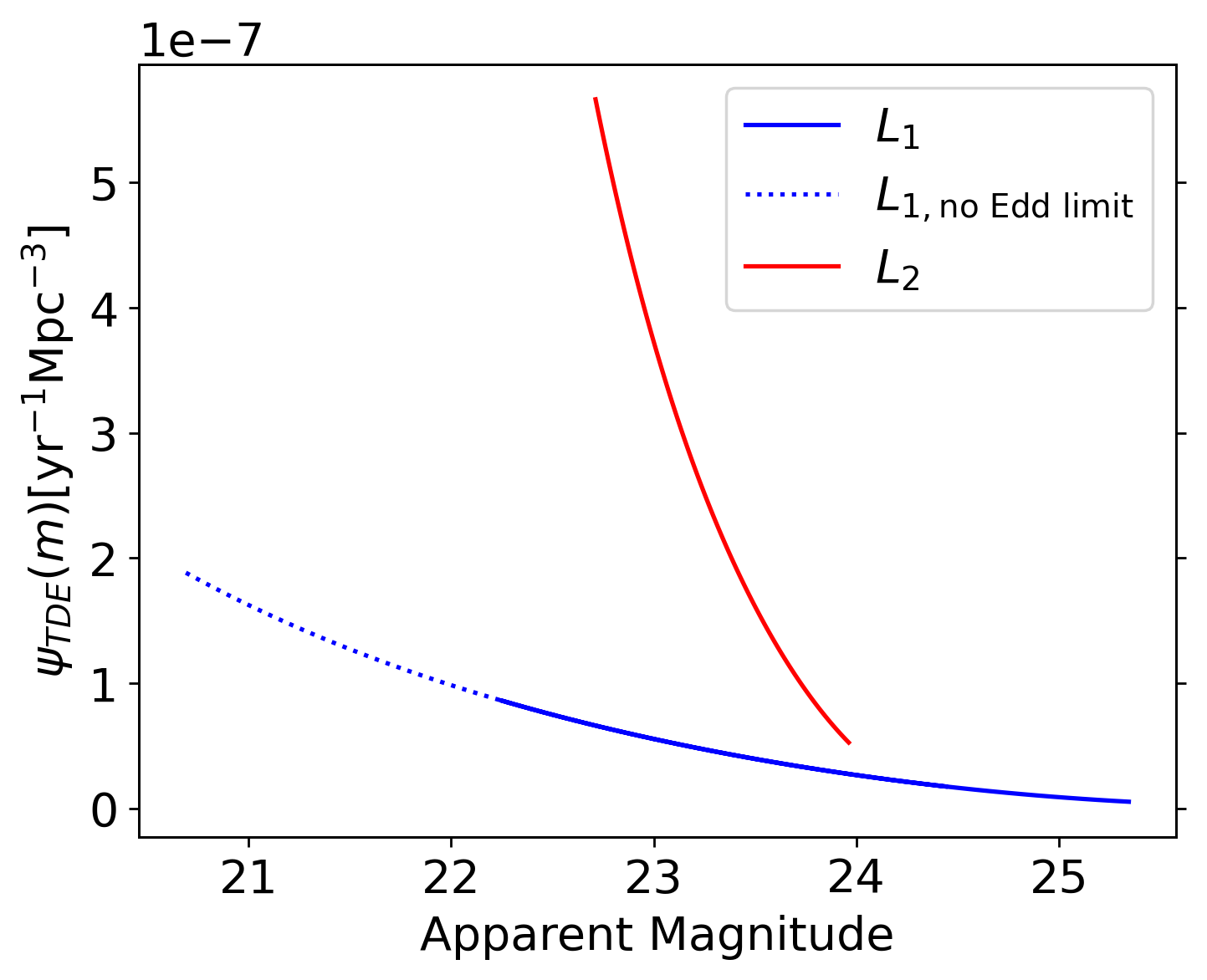}
\caption{Luminosity function in terms of magnitude for $L_{1}$ (the $L_{1}$ without an Eddington limit is given by the dotted blue line while the Eddington limited $L_{1}$ is shown in solid blue) and $L_{2}$ at a redshift of $z = 1.0$ where the apparent magnitude range shown corresponds to the $M_{\rm BH}$ range of $10^{5} - 10^{8}\,\Ms$ considered throughout this paper.
}
\label{fig:LF}
\end{figure}

Now that we have the expression for the apparent magnitude, we can determine ${\rm d}L/{\rm d}m_{\rm X}$. Since $\Xi$ does not depend on $L$, the derivative is simply,
\begin{align}
    \frac{\mathrm{d}L} {\mathrm{d}m_{X}}&= -\frac{\ln(10) L}{2.5}.
    \label{eq:dL_dmag}
\end{align}
The expression for $\psi_{\rm TDE}(m,z)$ can then be found using Equations (\ref{eq:mag}), (\ref{eq:xi}), and (\ref{eq:dL_dmag}) for $L_{1}$ and $L_{2}$ (See Appendix \ref{sec:lf_expression} for full expression). The resulting luminosity functions for $L_{1}$ and $L_{2}$ are shown in Figure \ref{fig:LF}.

Finally, using $\psi_{\rm TDE}(m)$, $z_{\rm max}$, $m_{\rm min}$ and $m_{\rm max}$, we can determine the unlensed TDE detection rates by computing Equation (\ref{eq:unlensed_rate}) for a given band.

\subsection{Unlensed rates of TDEs}
\label{subsec:unlensedTDEs:results}

We compute the unlensed detection rates for a range of magnitude cutoffs. This allows us to generalize the unlensed detection rates to more than LSST and ZTF by estimating these detection rates for surveys with any given survey area and magnitude limit.
The TDE magnitudes that we have considered in Section \ref{subsec:conversion2} are the brightness at the peak of the TDE light curve. Therefore, we consider magnitude cutoffs brighter than the limiting magnitudes of surveys in order to obtain more detections along the TDE light curve near their peak brightness to allow these TDEs to be properly detected and classified. We will refer to these cutoff magnitudes at peak as $m_\mathrm{lim,peak}$. 
Additionally, to have more realistic luminosities in the $L_1$ regime, we set an upper bound on $L_{1}$ to be the Eddington luminosity for the given $M_{\rm BH}$.

\begin{figure}[h!]
    \centering
\includegraphics[width = 8.cm]{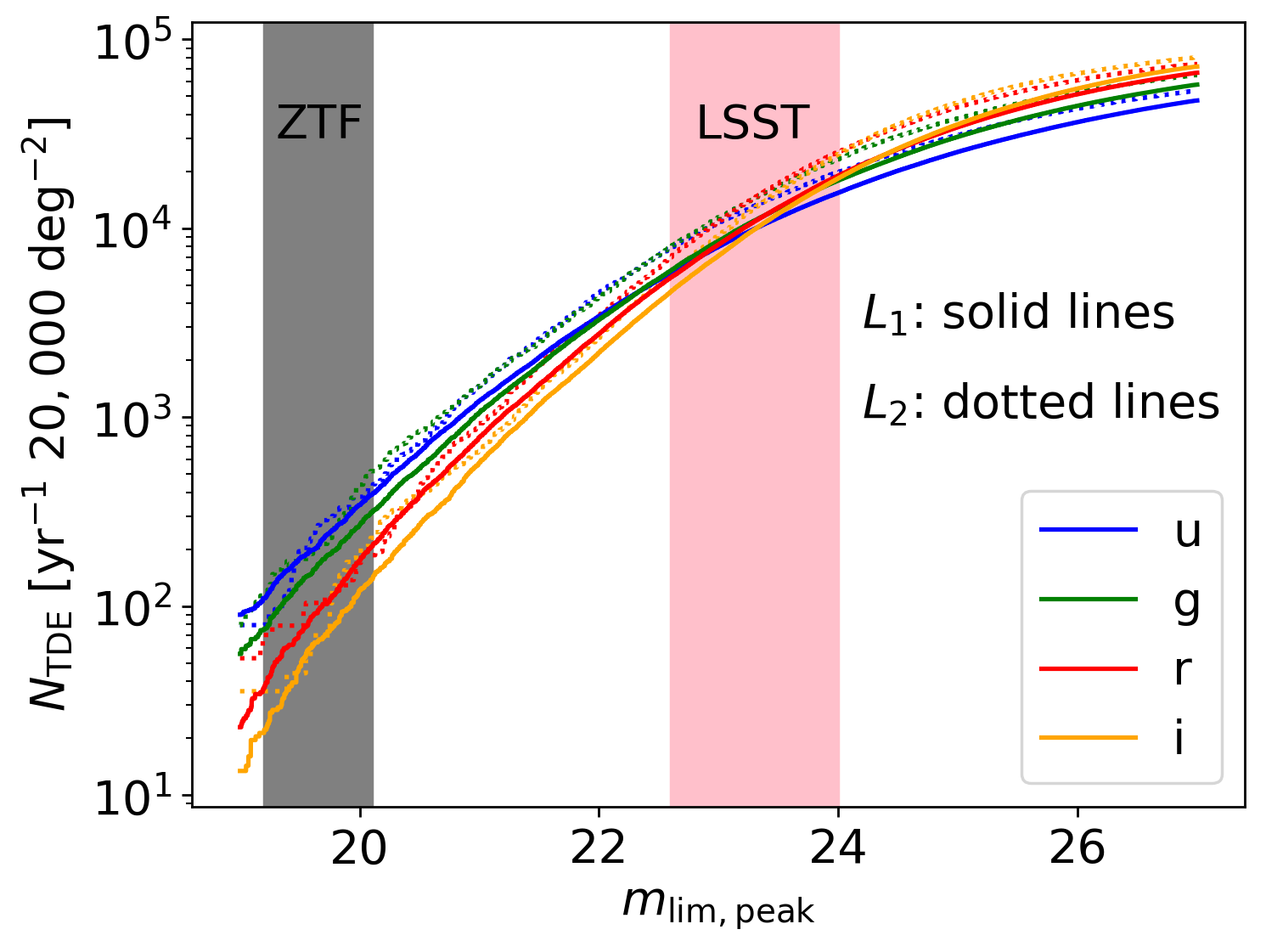}
\caption{The annual unlensed TDE detection rates as a function of the peak brightness, $m_\mathrm{lim,peak}$, at $T=2 \times 10^{4}$ K for $L_{1}$ in solid lines and for $L_{2}$ shown in dotted lines. 
The limiting magnitudes for the different filters of ZTF and LSST surveys are within the gray and pink shaded areas, respectively.  These limiting magnitudes are 0.7 brighter than the survey limiting magnitude $m_\mathrm{lim,survey}$ (i.e., $m_\mathrm{lim,peak} = m_\mathrm{lim,survey} - 0.7$), in order for TDEs to have multiple detections near their peak brightness above the limit $m_\mathrm{lim,survey}$.}
\label{fig:ntde_allM}
\end{figure}

In Figure \ref{fig:ntde_allM}, we show the annual unlensed detection rates for $L_{1}$ and $L_{2}$ at a temperature of $T = 2 \times 10^{4}$ K to illustrate the unlensed detection rates assuming a generic $m_\mathrm{lim,peak}$, but as we will soon see this temperature matches well with observations. We also provide the LSST and ZTF detection regions to indicate where these magnitude ranges fall within the curves. This LSST model, denoted LSST1, defines the magnitude cutoff as a magnitude value of 0.7 less than the LSST magnitude band limit, $m_\mathrm{lim,survey}$ \citep{Oguri_2010}. We apply the same cutoff criterion to the ZTF limit as in LSST1. The $m_\mathrm{lim,survey}$ values for LSST and ZTF are  given by ($u$, $g$, $r$, $i$) = (23.3, 24.7, 24.3, 23.7) \citep[e.g.,][]{Huber_2021,Lochner_2022} and ($g$, $r$, $i$) = (20.8, 20.6, 19.9) \citep{Bellm_2018}, respectively, where ZTF does not observe in the $u$ band. 

\begin{figure*}[h!]
    \centering
\includegraphics[width = 9cm]{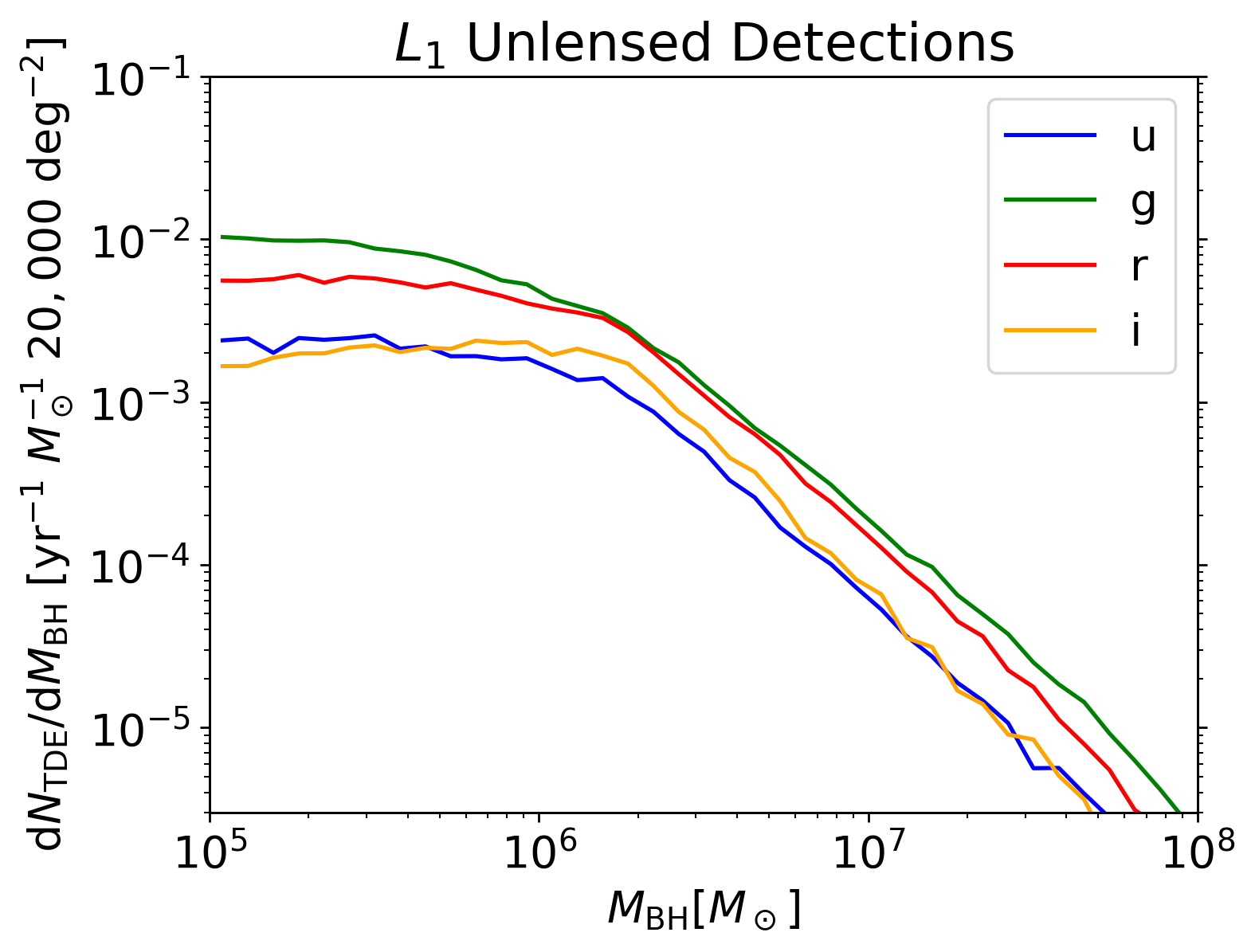}
\includegraphics[width = 9cm]{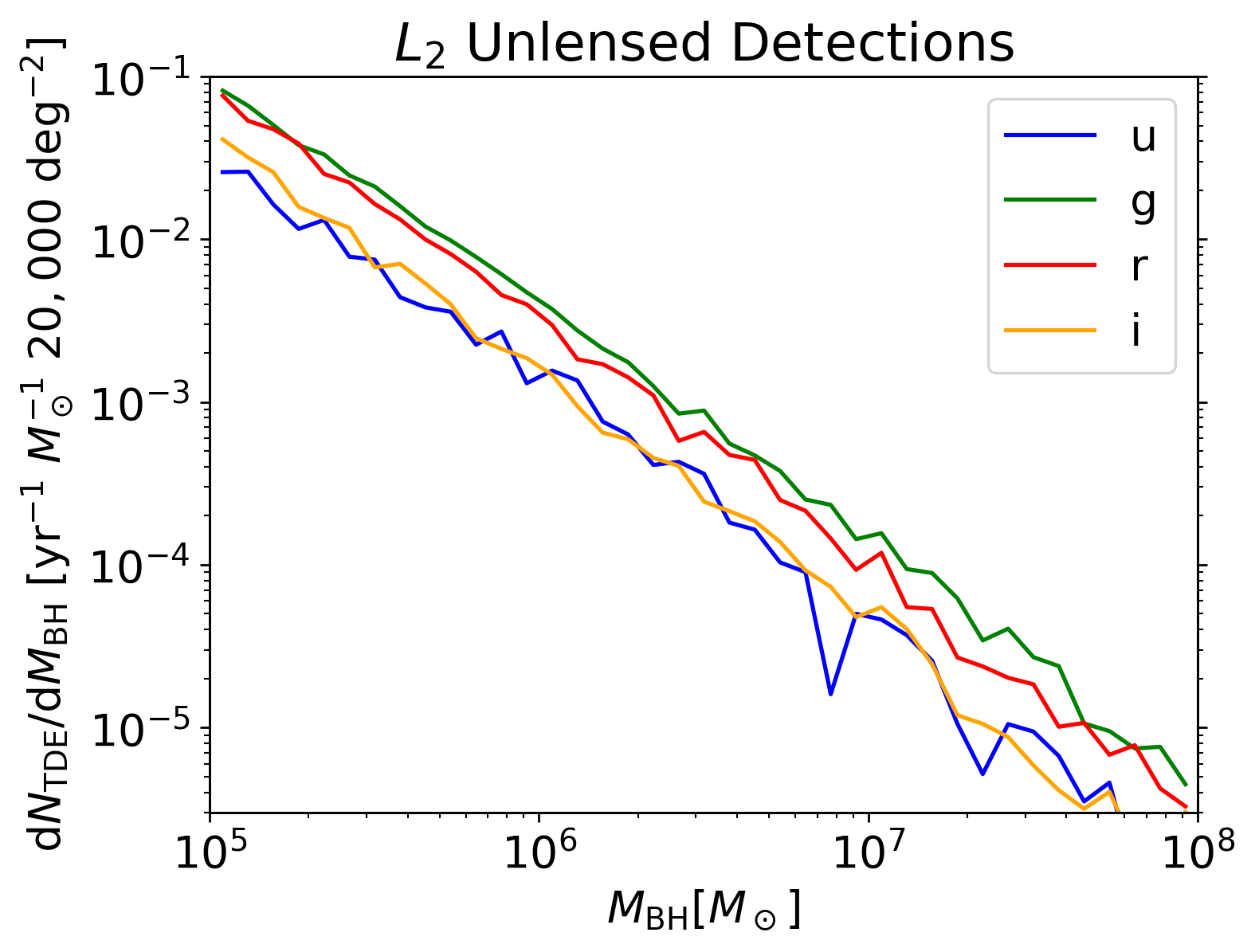}
\caption{The annual unlensed $L_{1}$ (\textit{left} panel) and $L_{2}$ (\textit{right} panel) detection rates for $u$, $g$, $r$ and $i$ bands as a function of $M_\mathrm{BH}$ assuming LSST1 limiting magnitudes.}
\label{fig:unlensedrates_mbh}
\end{figure*}

From Figure \ref{fig:ntde_allM}, we see that as $m_\mathrm{lim,peak}$ increases, the annual unlensed detection rates for both luminosities increase correspondingly as observing at these fainter magnitudes enables the detections of observations at further redshifts. $L_{1}$ and $L_{2}$ have comparable detection rates with $L_{2}$ producing slightly larger rates in general.
We are also able to observe how the more successful band changes with the peak magnitude. For a brighter $m_\mathrm{lim,peak}$, the $u$ and $g$ bands produce the greater number of detections, but both $L_{1}$ and $L_{2}$ experience a crossing at  $m_\mathrm{lim,peak} \simeq 23.3$. At fainter magnitudes, the $i$ and $r$ bands produce more detections which is a result of these sources being at further distances, and this implies that their observed fluxes would be redshifted. The relation between an increasing redshift and the observed flux is seen in Figure \ref{fig:l2_flux} when comparing the ZTF-like detections, at $z=0.2$, and the LSST-like detections, at $z=1.0$. At all temperatures, when the source is at a further redshift, this peak of the observed flux is pushed to longer wavelengths which indicates that at such larger redshifts we expect $i$ and $r$ bands to dominate the detections.

We also present the unlensed detection rates as a function of $M_\mathrm{BH}$ for the LSST1 survey in Figure \ref{fig:unlensedrates_mbh}. We find that the lower BH masses dominate the detections and that the detections decrease with BH mass.
The distribution shape differs notably between $L_{1}$ and $L_{2}$ for lower BH masses because the luminosity of $L_{1}$ is Eddington limited, and by Equation \ref{eq:L1_simplified} this significantly impacts the lower masses. Thus, limiting $L_{1}$ luminosity flattens off the distribution at lower masses, especially for $M_\mathrm{BH} < 10^{6} \Ms$. 

\begin{figure}[h!]
    \centering
\includegraphics[width = 8.6cm]{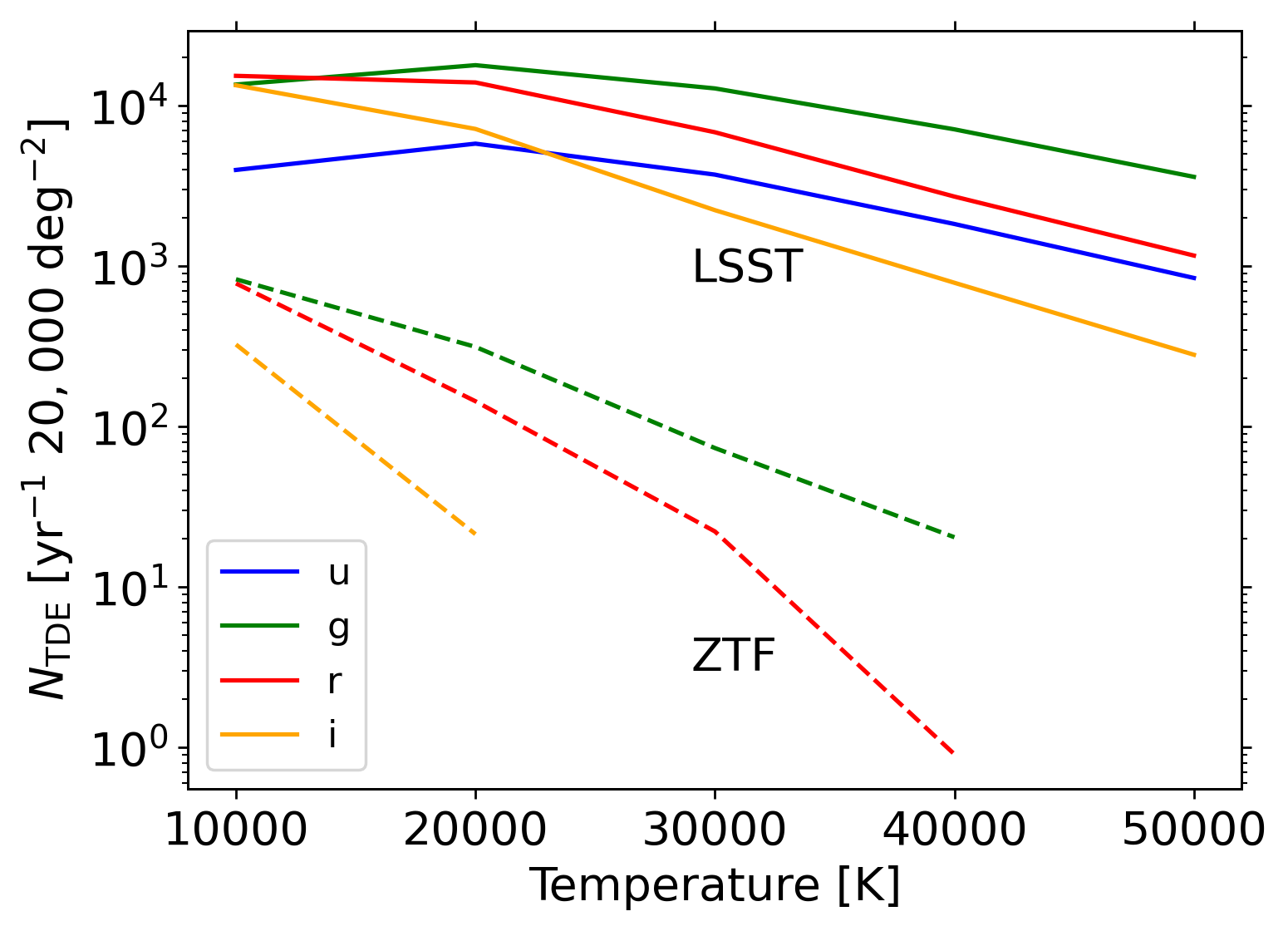}
\caption{The $L_{1}$ annual TDE detection rate with varying temperatures for LSST1, shown in solid lines, and ZTF, shown in dashed lines.}
\label{fig:l1_ntde_allT}
\end{figure}

\begin{figure}[h!]
    \centering
\includegraphics[width = 8.6cm]{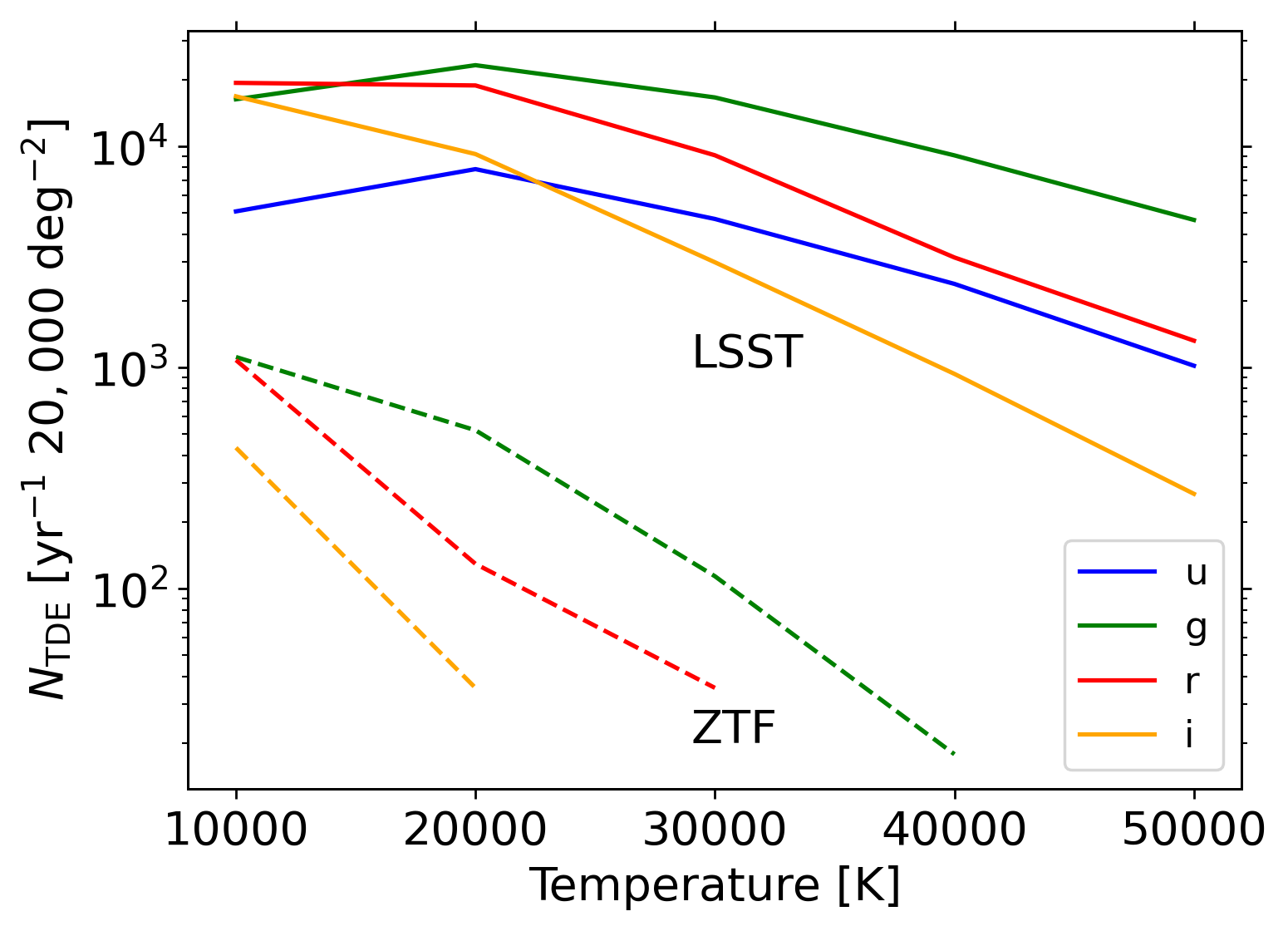}
\caption{The $L_{2}$ annual TDE detection rate with varying temperatures for LSST1, shown in solid lines, and ZTF, shown in dashed lines.}
\label{fig:l2_ntde_allT}
\end{figure}

Next, we calculate the unlensed TDE detection rates using Equation (\ref{eq:unlensed_rate}) as a function of temperature for the models referenced previously, for LSST1 and ZTF. In Figures \ref{fig:l1_ntde_allT} and \ref{fig:l2_ntde_allT}, we show that LSST1 results in higher detections than that of  ZTF since it employs fainter peak magnitude limits. We can see that the $g$ band detection rates are generally higher than those for other bands within the temperature range considered, independent of the assumption for the luminosity. Figure \ref{fig:l1_ntde_allT} shows that the $L_{1}$ $g$ band detection rate for LSST1 is $1.4\times 10^{4}$ at $T=1 \times 10^{4}$ K which decreases to $3.6 \times 10^{3}$ at $T=5 \times 10^{4}$ K. Whereas, ZTF's largest detection rate is $8.3\times 10^{2}$ for $L_{1}$ $g$ band at $T=1 \times 10^{4}$ K before the rates proceed to decline. The trends for the unlensed rates given varying temperatures persist for both $L_{1}$ and $L_{2}$.

\begin{table*}
\centering
\begin{tabular}{  c c} 
\hline
Observational survey & $m_\mathrm{lim,peak}$ \\
\hline
LSST1 & $m_\mathrm{lim,survey} - 0.7$ \\
LSST2 & $m_\mathrm{lim,survey} - 2.0$ \\
ZTF & $m_\mathrm{lim,survey} - 0.7$ \\
\hline
\end{tabular}
\caption{The three observational magnitude cutoffs used for computing the unlensed TDE detection rates. The survey magnitude limit for each band $-$ 0.7 is following \citet{Oguri_2010}. The survey magnitude limit for each band $-$ 2.0 is following \citet{Bricman_2020}. The $m_\mathrm{lim,survey}$ values for LSST and ZTF are ($u$, $g$, $r$, $i$) = (23.3, 24.7, 24.3, 23.7) \citep[e.g.,][]{Huber_2021, Lochner_2022} and ($g$, $r$, $i$) = (20.8, 20.6, 19.9) \citep{Bellm_2018}, respectively.}\label{tab:survey_methods}
\end{table*}

\begin{table*}
\centering
\begin{tabular}{  c c c c c c} 
\hline
Temperature & Observational survey & $u$ [$\times 10^{3}$] & $g$ [$\times 10^{3}$] & $r$ [$\times 10^{3}$] & $i$ [$\times 10^{3}$]\\
\hline
\hline
1 $\times 10^{4}$K & LSST1 & 4.0 & 14 & 15 & 13 \\
1 $\times 10^{4}$K & LSST2 & 1.6 & 6.8 & 7.1 & 5.1 \\
1 $\times 10^{4}$K & ZTF & - & 0.83 & 0.78 &0.32 \\
\hline
2 $\times 10^{4}$K & LSST1 & 5.8 & 18 & 14 &7.2 \\
2 $\times 10^{4}$K & LSST2 & 1.7 & 6.6 & 4.0 & 1.5 \\
2 $\times 10^{4}$K & ZTF & - & 0.31 & 0.14 & 0.021 \\
\hline
3 $\times 10^{4}$K & LSST1 & 3.7 & 13 & 6.8 & 2.2 \\
3 $\times 10^{4}$K & LSST2 & 0.80 & 3.2& 1.3 & 0.34 \\
3 $\times 10^{4}$K & ZTF & - & 0.073 & 0.022 & 0.0 \\
\hline
4 $\times 10^{4}$K & LSST1 & 1.8 & 7.1 & 2.7 & 0.79 \\
4 $\times 10^{4}$K & LSST2 & 0.30 & 1.4 & 0.43 & 0.091 \\
4 $\times 10^{4}$K & ZTF & - & 0.020 & 0.00091 & 0.0 \\
\hline
5 $\times 10^{4}$K & LSST1 & 0.84 & 3.6 & 1.2 & 0.28 \\
5 $\times 10^{4}$K & LSST2 & 0.12 & 0.55 & 0.15 & 0.027 \\
5 $\times 10^{4}$K & ZTF & - & 0.00 & 0.00 & 0.0 \\
\hline
\end{tabular}
\caption{Annual unlensed TDE detection rates for $L_{1}$ given different temperatures for the three surveys' limiting magnitudes in Table \ref{tab:survey_methods} assuming a survey area of 20,000 $\mathrm{deg^{2}}$ for all surveys.} 
\label{tab:unlensedrate_l1}
\end{table*}

\begin{table*}
\centering
\begin{tabular}{  c c c c c c} 
\hline
Temperature & Observational survey & $u$ [$\times 10^{3}$] & $g$ [$\times 10^{3}$] & $r$ [$\times 10^{3}$] & $i$ [$\times 10^{3}$]\\
\hline
\hline
1 $\times 10^{4}$K & LSST1 & 5.1 & 16 & 19 & 17 \\
1 $\times 10^{4}$K & LSST2 & 1.9 & 8.5 & 9.0 & 6.4 \\
1 $\times 10^{4}$K & ZTF & - & 1.1 & 1.1 & 0.43 \\
\hline
2 $\times 10^{4}$K & LSST1 & 7.9 & 23 & 19 & 9.2 \\
2 $\times 10^{4}$K & LSST2 & 2.1 & 8.8 & 4.9 & 1.8 \\
2 $\times 10^{4}$K & ZTF & - & 0.52 & 0.13 & 0.035 \\
\hline
3 $\times 10^{4}$K & LSST1 & 4.7 & 17 & 9.1 & 3.0 \\
3 $\times 10^{4}$K & LSST2 & 1.0 & 3.9 & 1.6 & 0.48 \\
3 $\times 10^{4}$K & ZTF & - & 0.11 & 0.035 & 0.0 \\
\hline
4 $\times 10^{4}$K & LSST1 & 2.4 & 9.1 & 3.1 & 0.93 \\
4 $\times 10^{4}$K & LSST2 & 0.35 & 1.7 & 0.49 & 0.12 \\
4 $\times 10^{4}$K & ZTF & - & 0.018 & 0.0 &0.0 \\
\hline
5 $\times 10^{4}$K & LSST1 & 1.0 & 4.6 & 1.3 & 0.27 \\
5 $\times 10^{4}$K & LSST2 & 0.13 & 0.79 & 0.20 & 0.018 \\
5 $\times 10^{4}$K & ZTF & - & 0.0 & 0.0 & 0.0\\
\hline
\end{tabular}
\caption{Same as Table~\ref{tab:unlensedrate_l1}, but for $L_{2}$.}
\label{tab:unlensedrate_l2}
\end{table*}

\begin{figure}[h!]
    \centering
\includegraphics[width = 8.cm]{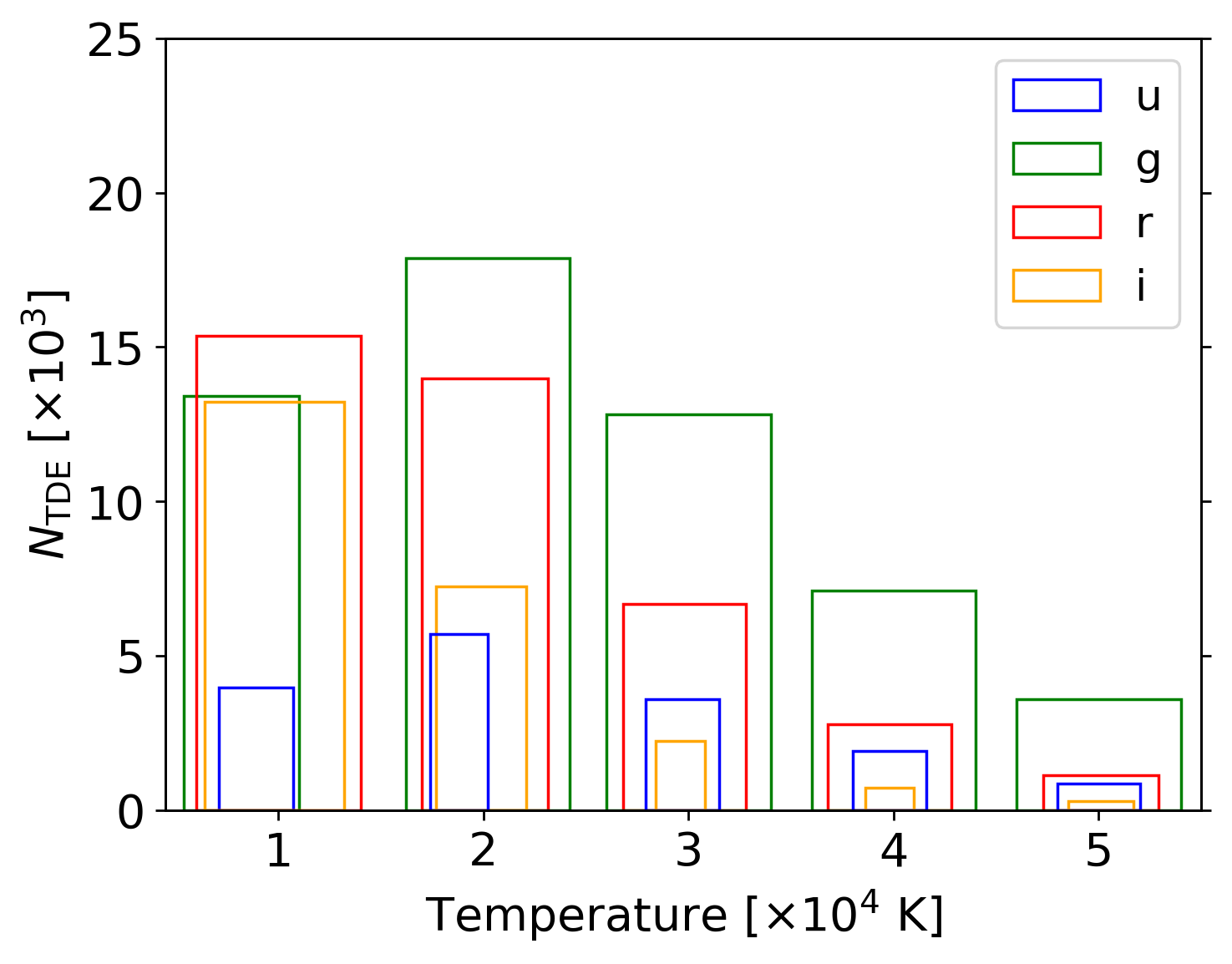}
\caption{The number and hierarchy of unlensed $L_{1}$ detections for each of the four bands across the five temperatures we consider given LSST1 magnitude cutoffs. The band that results in the most detections for each temperature is the tallest rectangle (i.e., the $r$ band for $T=1 \times 10^{4}$ K and $g$ band for all other temperatures) where the height of the rectangle indicates the number of detections, and a rectangle contained within another implies that a detection in that band is also detected in the encompassing band. For instance, in the $T=3 \times 10^{4}$ K case, the yellow rectangle for $i$ band is contained entirely within the green, red and blue rectangles for $g$, $r$ and $u$ bands, respectively. This indicates that all $i$ band detections are also detected within $g$, $r$ and $u$ bands. The partial overlapping for $T=1 \times 10^{4}$ K ($g$ and $i$) and $2\times 10^{4}$ K ($u$ and $i$) means those detections do not establish a full hierarchy. Note that the width of the rectangles does not convey any significant meaning.}
\label{fig:l1_unlensed_hierarchy}
\end{figure}

\begin{figure}[h!]
    \centering
\includegraphics[width = 8.cm]{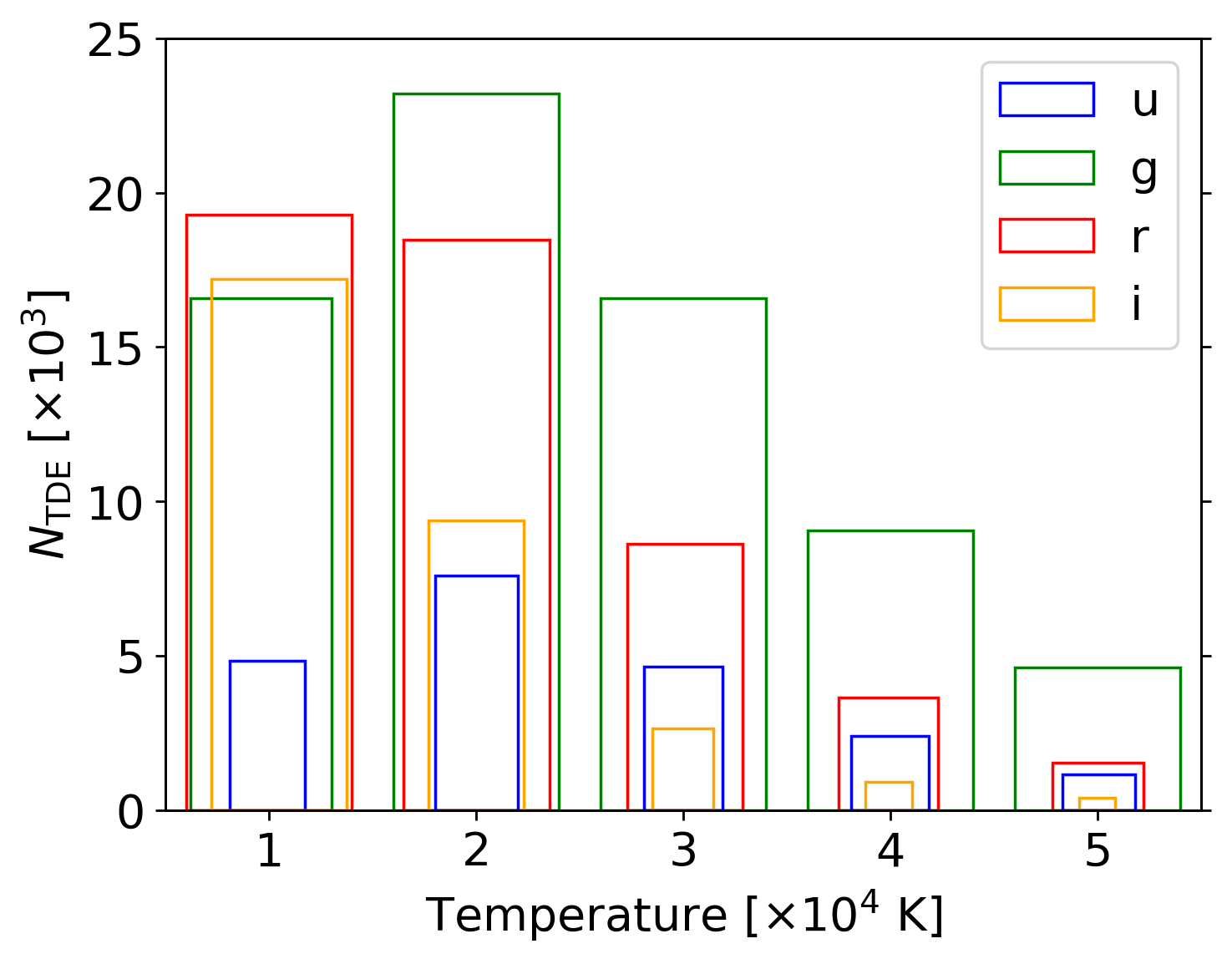}
\caption{The number and hierarchy of unlensed $L_{2}$ detections for each of the four bands across the five temperatures we consider given LSST1 magnitude cutoffs. The band that results in the most detections for each temperature is the tallest rectangle where the height of the rectangle indicates the number of detections, and a rectangle contained within another implies that a detection in that band is also detected in the encompassing band. Note that the width of the rectangles does not convey any significant meaning.}
\label{fig:l2_unlensed_hierarchy}
\end{figure}

In Figures \ref{fig:l1_unlensed_hierarchy} and \ref{fig:l2_unlensed_hierarchy}, we present the number of unlensed TDE detections per band (given by the height of the histograms) and the hierarchy of the four bands we consider for the unlensed $L_{1}$ and $L_{2}$ detections, respectively, in the range of $T= (1 - 5) \times 10^{4}$ K assuming LSST1 magnitude cutoffs. We obtain this hierarchy by computing how many detections within a particular band are also detected in the remaining three bands, and this allows us to determine the overlap of detections across the bands. For $L_{1}$ in Figure \ref{fig:l1_unlensed_hierarchy}, we see that there is a complete hierarchy for $T=(3 - 5) \times 10^{4}$ K; at these temperatures, we find that $g \supset r \supset u \supset i$ where $a \supset b$ means $a$ encompasses $b$. For the remaining two temperatures, there are partial overlaps between bands rather than each successive band's detections being encompassed in those of the previous band.
%the previous detections. 
For example, at $T=1 \times 10^{4}$ K, the green $g$ histogram has a section that is outside both the $r$ and $i$ histograms, so this indicates that there are detections in $g$ band that are not seen in either $r$ or $i$ bands. There is a similar occurrence for the $u$ band in comparison to the $i$ band at $T=2 \times 10^{4}$ K. Partial overlappings are likely due to a TDE's magnitude being fainter than a band's $m_\mathrm{lim,peak}$ at LSST1 magnitude cutoffs. As an example, we show in Appendix \ref{sec:appendix_mag_conversion} the relation between $m_\mathrm{r}$ and $m_\mathrm{g}$ at two different temperatures for three constant luminosity cases, and it can be seen that some luminosities will be detected in $r$ and not in $g$ and vice versa depending on the particular $L$ and $T$. These relations exist across all bands and can help to explain why TDE detections may be missed in the other bands.

For $L_{2}$ in Figure \ref{fig:l2_unlensed_hierarchy}, we observe that there is a complete hierarchy for $T \geq 2 \times 10^{4}$ K. For example, for $T=2\times10^{4}$ K, $g \supset r \supset i \supset u$. For the $T= 1 \times 10^{4}$ K case, we observe that $r$ band is the dominating filter that encompasses all other detections. However, while $i$ band detects more TDEs than $g$ band, $i$ band does not detect all of the $g$ band TDEs, but despite this all of the $u$ band detections are also seen in both $g$ and $i$ which is why the blue $u$ rectangle is contained entirely within the $g$ and $i$ band overlapping regions.
However, we note that the number of $g$ band detections missed in $i$ band is approximately on the order of the statistical fluctuations. The number of detections in each of the bands and the overlap between bands are also tabulated in Appendix \ref{sec:appendix_mag_conversion}.

Next, we compute these unlensed detection rates for an additional LSST model. This second LSST model, LSST2, defines the cutoff as 2.0 less than that of the LSST $m_\mathrm{lim,survey}$ \citep{Bricman_2020}. We summarize the three magnitude cutoffs we utilize in Table \ref{tab:survey_methods}. We provide these unlensed detection rates as a function of TDE temperature in Tables \ref{tab:unlensedrate_l1} and \ref{tab:unlensedrate_l2} for all models. From both Tables, it is evident that LSST1 results in the highest detections. It then follows that, as the more conservative LSST approach, LSST2 detects fewer TDEs than LSST1 but still greater than that of ZTF. For $L_{1}$, the LSST1 detection rates at $T = 1 \times 10^{4}$ K are $1.5 \times 10^{4}$, $1.3\times 10^{4}$ and $1.4\times 10^{4}$ for $r$, $i$, and $g$ bands, respectively, and these detection rates are $1.9\times 10^{4}$, $1.7\times 10^{4}$ and $1.6\times 10^{4}$ for $L_{2}$. As temperature increases, $g$ band results in the highest detections. The $g$ band detections peak at $2 \times 10^{4}$ K for both luminosities with a detection rate of $1.8\times 10^{4}$ for $L_{1}$ and $2.3 \times 10^{4}$ for $L_{2}$. Then, as the temperature increases further, the detections decrease because the flux peak shifts more into the ultraviolet range as seen in Figure \ref{fig:l2_flux}.

Our unlensed detection rates are generally higher by a few to $20$ than the ZTF detection rates in the $r$ and $g$ bands, which are on the order of 10 - 20 TDEs annually \citep[see, e.g., ][]{2023arXiv230306523Y} with $T\simeq (2 - 3)\times 10^{4}$ K \citep[e.g.,][]{Hammerstein+2023}. We attribute this difference to a possible stronger contribution of low-mass $M_{\rm BH}$ to the rate in our estimates (especially for $L_{2}$) and the incompleteness of observations. While the inference of $M_{\rm BH}$ from observations is highly dependent on the emission models, the inferred $M_{\rm BH}$ for ZTF TDEs tends to be greater than $10^{6}$ M$_{\odot}$ \citep[e.g.,][]{Hammerstein+2023}.

\section{Rates of lensed TDEs}
\label{sec: Rates of lensed TDE}
\subsection{Overview of method}
We utilize the code developed by \cite{Oguri_2010} that was used to produce mock catalogs for strongly lensed supernovae and quasars. We implemented the rate calculation of unlensed TDEs into the code to produce similar mock catalogs for strongly lensed TDEs. This method constructs a source table given a magnitude range, redshift range and the luminosity function of the source that was obtained in Section \ref{subsec:conversion2} and shown in Figure \ref{fig:LF}.
Here, we set the magnitude to span the range from a sufficiently bright magnitude ($m = 15.0$) to a magnitude of $m_\mathrm{lim,peak} = 27.0$. From the output catalog of the lensed TDEs, we can subsequently apply brighter limiting magnitude cutoffs such that we can obtain the detection rates for any cadenced imaging survey with single-epoch limiting magnitude brighter than $m_\mathrm{lim,peak}=27.0$.
We probe from the local universe to a sufficiently high redshift of $z = 6.8$ for $L_{1}$ and $z = 5.5$ for $L_{2}$ where these values are chosen to be greater than the largest $z_\mathrm{max}$ across all bands for both cases to avoid missing any detections by not probing to far enough distances. 

In this method, the lenses are taken to be elliptical galaxies with mass distributions described by the singular isothermal ellipsoid. The environment effects of each lens are characterized by an external shear. While \citet{Oguri_2010} assumes no redshift evolution for the velocity dispersion function, in this paper the lenses are distributed such that their velocity dispersion functions evolve with redshift as defined by Equation (11) in \citet{Oguri_2018}, and we set a maximum redshift for the lenses at $z = 2.0$.
The lensing probability of a given source is computed, and from this the total number of strong lenses can be determined. The lensing for that source is then simulated.
For a strongly lensed double system to be detected, we require that both images must have a magnitude brighter than $m_\mathrm{lim,peak}$; for a quadruple system, we require that three images must be brighter than $m_\mathrm{lim,peak}$.
The choice of the third brightest image is due to the fact that this image tends to appear first, and it is often further from the brightest image which allows it to be more easily identified as a strongly lensed system.
We then obtain the image separation, magnitude of the lensed third brightest image (or fainter image for double systems), redshifts of the lens and TDE as well as the locations and magnifications of all images within the lensed systems.

We use the TDE luminosity function in terms of magnitude, $\psi_{\rm TDE}(m,z)$, to construct the TDE source table. To reduce statistical noise (i.e., statistical fluctuations given the typically low rates of lensed TDEs) and obtain a sufficiently large number of mock lenses to produce distributions, we oversample each simulation run by a factor of 1000 and then renormalize the output number of lenses by this factor. We limit the $L_{1}$ luminosity to the Eddington luminosity, as for the unlensed case. The luminosity is then used to compute the unlensed TDE magnitude and together with the magnification it is used to compute the lensed magnitude.

\subsection{Results}
We compute the strongly lensed TDE detection rates for $L_{1}$ and $L_{2}$ assuming a constant temperature of $2 \times 10^{4}$ K.  

In a strongly lensed system, the image separation must be large enough to be resolved, so we retain only detections with image separations within the range of $0.5\arcsec$ - $4\arcsec$. 
The choice of the image separation cut off at $4\arcsec$ is because we focus on galaxy-scale lenses in this study whose image separations are typically $\lesssim 4\arcsec$ (larger image separations would correspond to galaxy groups or galaxy clusters as lenses).
We show the strongly lensed detection rates per effective year for various bands assuming a variable $m_\mathrm{lim,peak}$ for each of the $L_1$ and $L_2$ luminosity functions.

\subsubsection{Rates}

\begin{figure*}[h!]
    \centering
\includegraphics[width = 9cm]{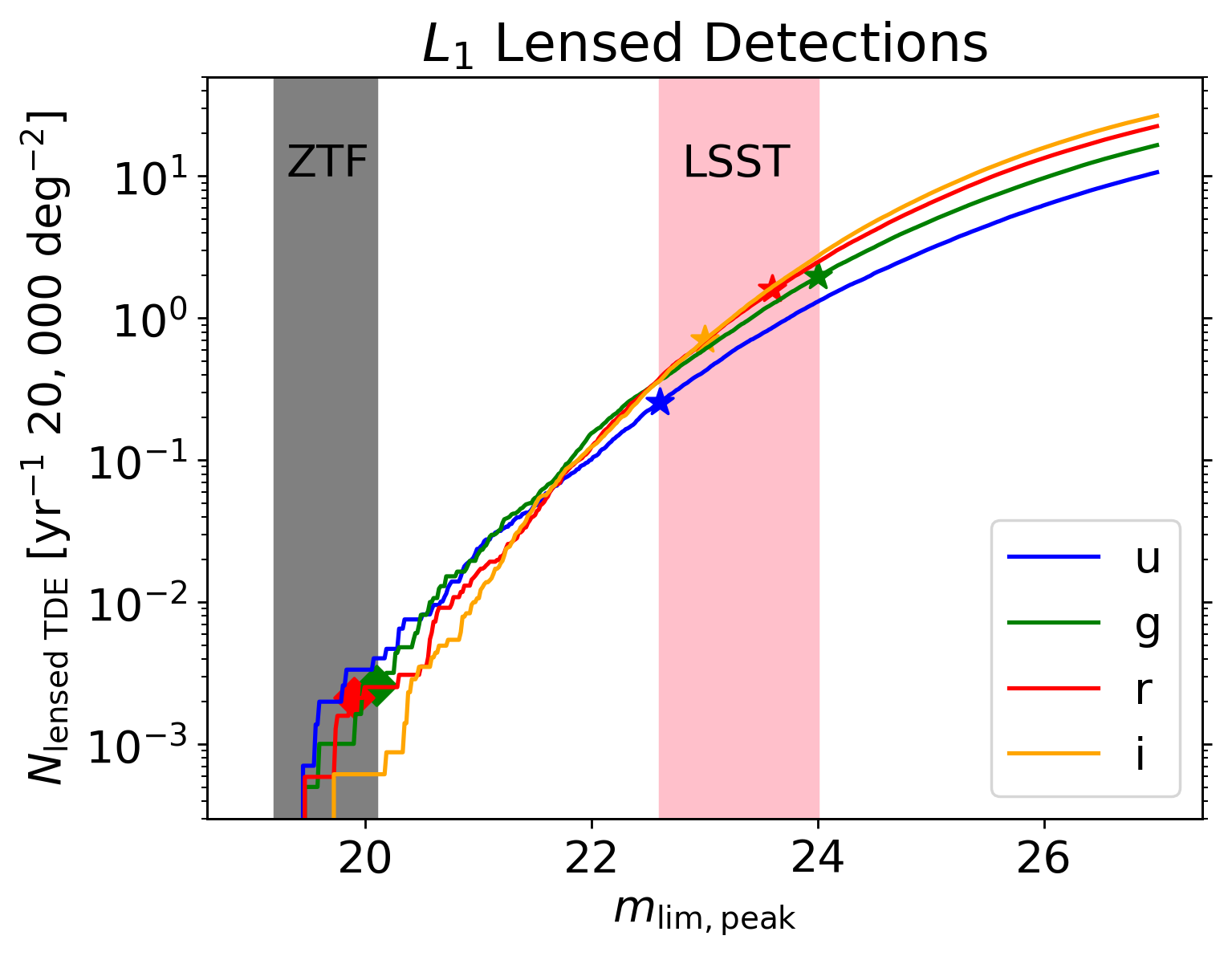}
\includegraphics[width = 9cm]{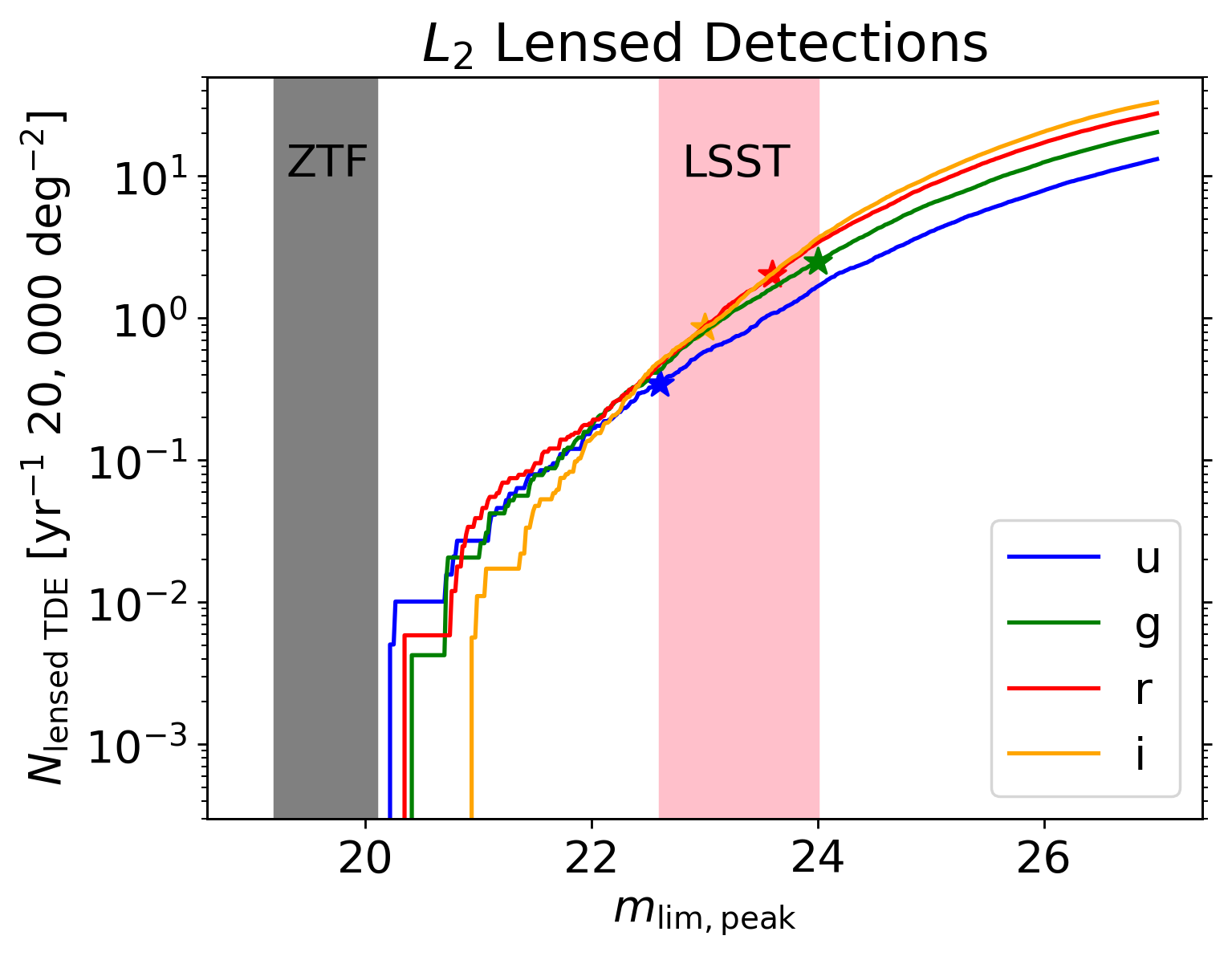}
\caption{The strongly lensed $L_{1}$ (\textit{left} panel) and $L_{2}$ (\textit{right} panel) detection rates per effective year of observation for $u$, $g$, $r$ and $i$ bands assuming a variable peak magnitude, $m_\mathrm{lim,peak}$. The $m_\mathrm{lim,peak}$ ranges for ZTF and LSST1 assuming a survey area of 20,000 $\mathrm{deg^{2}}$ are shown in gray and pink, respectively. The LSST1 $m_\mathrm{lim,peak}$ values for each of the four bands are overlaid along the curves in stars, and the ZTF $m_\mathrm{lim,peak}$ values for $g$, $r$ and $i$ are overlaid along the curves in diamonds.}
\label{fig:lensed_rates}
\end{figure*}

As $m_\mathrm{lim,peak}$ increases, the strongly lensed detection rates increase as well given that we are able to detect fainter and fainter lensed TDEs which, in turn, implies that we are probing at further and further distances. The lensed detection rates per effective year of observation for each of the four bands
along with the LSST1 and ZTF magnitude cutoffs are shown in Figure \ref{fig:lensed_rates} for $L_{1}$ (left panel) and $L_{2}$ (right panel). These rates assume a survey area of 20,000 $\mathrm{deg}^{2}$, and since the area of ZTF is larger than LSST, these rates become a lower limit estimate for ZTF. From this, it is unlikely that ZTF will detect a strongly lensed TDE. However, the results are more promising for that of LSST. Since theses detection rates are for a single effective year of observation, these results must be multiplied by the effective duration of LSST to determine the detection rates for the entirety of LSST. The lensed detection rates for the duration of LSST for $L_{1}$ and $L_{2}$ are shown in Table \ref{tab:lensed_rates} assuming an effective duration of 4.5 years which is typical for the LSST baseline observing strategy \citep{Huber_2019,Lochner_2022}. 

Specifically, for $L_{1}$, the lensed detection rates per effective year of observation are ($u$, $g$, $r$, $i$) = (11, 17, 22, 27) at $m_\mathrm{lim,peak}=27.0$. The lensed rates for $L_{2}$ are ($u$, $g$, $r$, $i$) = (13, 20, 28, 33).
The lensed LSST detection rates for $L_{1}$ and $L_{2}$ show that $g$ band will be the most successful (as shown by the star symbols in Figure \ref{fig:lensed_rates}). At further magnitudes, the trend resembles that of the unlensed case with the $i$ band resulting in the highest detections for both $L_{1}$ and $L_{2}$ as shown in Figure \ref{fig:lensed_rates}. Thus, at these fainter peak magnitudes, $i$ and $r$ bands are expected to produce the most strongly lensed detections for both luminosities. With $m_\mathrm{lim,peak}=27.0$ for an effective duration of 4.5 years, $L_{2}$ is expected to detect the most lensed TDEs within the $i$ band; given LSST1 cutoffs, the $r$ band is expected to detect approximately 5 more lensed TDEs than $i$ band, but at this further $m_\mathrm{lim,peak}=27.0$, $i$ band is anticipated to detect 23 more lensed TDEs than $r$ band. 
$L_{1}$ is projected to detect on the order of 50 more strongly lensed TDEs in $i$ band than in $g$ band at $m_\mathrm{lim,peak}=27.0$.

Given the lensed detection rates, we can further investigate the occurrence of a lensed detection by determining the ratio of lensed to unlensed detections. These ratios assuming an LSST1 survey for all bands of $L_{1}$ are ($u$, $g$, $r$, $i$) = (4.2, 11, 8.3, 9.9) $\times 10^{-5}$, and similarly for $L_{2}$ these ratios are ($u$, $g$, $r$, $i$) = (4.2, 11, 11, 9.2) $\times 10^{-5}$.
Thus, $L_{1}$ and $L_{2}$ produce similar relative ratios of lensed to unlensed events, especially for $u$ and $g$ bands. Additionally, these ratios indicate that for every $10^{4}$ unlensed TDE detections, we may detect ${\sim}1$ lensed event. Although the scaling factor would depend on several assumptions, it can serve as a rough but useful guide to estimate the lensed rates from the unlensed rates without detailed calculations.

\begin{table*}
\centering
\begin{tabular}{  c c c c c c} 
\hline
Luminosity Model & Observational Survey & $u$ & $g$ & $r$ & $i$ \\
\hline
$L_{1}$ & LSST1 & 1.1 & 8.7 & 7.1 & 3.2 \\
$L_{2}$ & LSST1 & 1.5 & 11 & 9.1 & 3.8 \\
\hline
\end{tabular}
\caption{Lensed TDE detection rates for $L_{1}$ and $L_{2}$ using the LSST1 magnitude cutoffs at $T=2 \times 10^{4}$ K for the entire duration of LSST assuming a typical effective duration of 4.5 years. These lens systems have multiple TDE images separated by $>0.5\arcsec$.}
\label{tab:lensed_rates}
\end{table*}

\subsubsection{Lens properties}
\label{sec:lens properties}
The distributions showing the lensing properties are constructed using the full catalog that was oversampled by a factor of 1000 for each band, and the resulting figures are then normalized by this factor to show the lensed detection rates for a single effective year of observation.

\begin{figure*}[ht]
\centering
{\includegraphics[width=6cm]{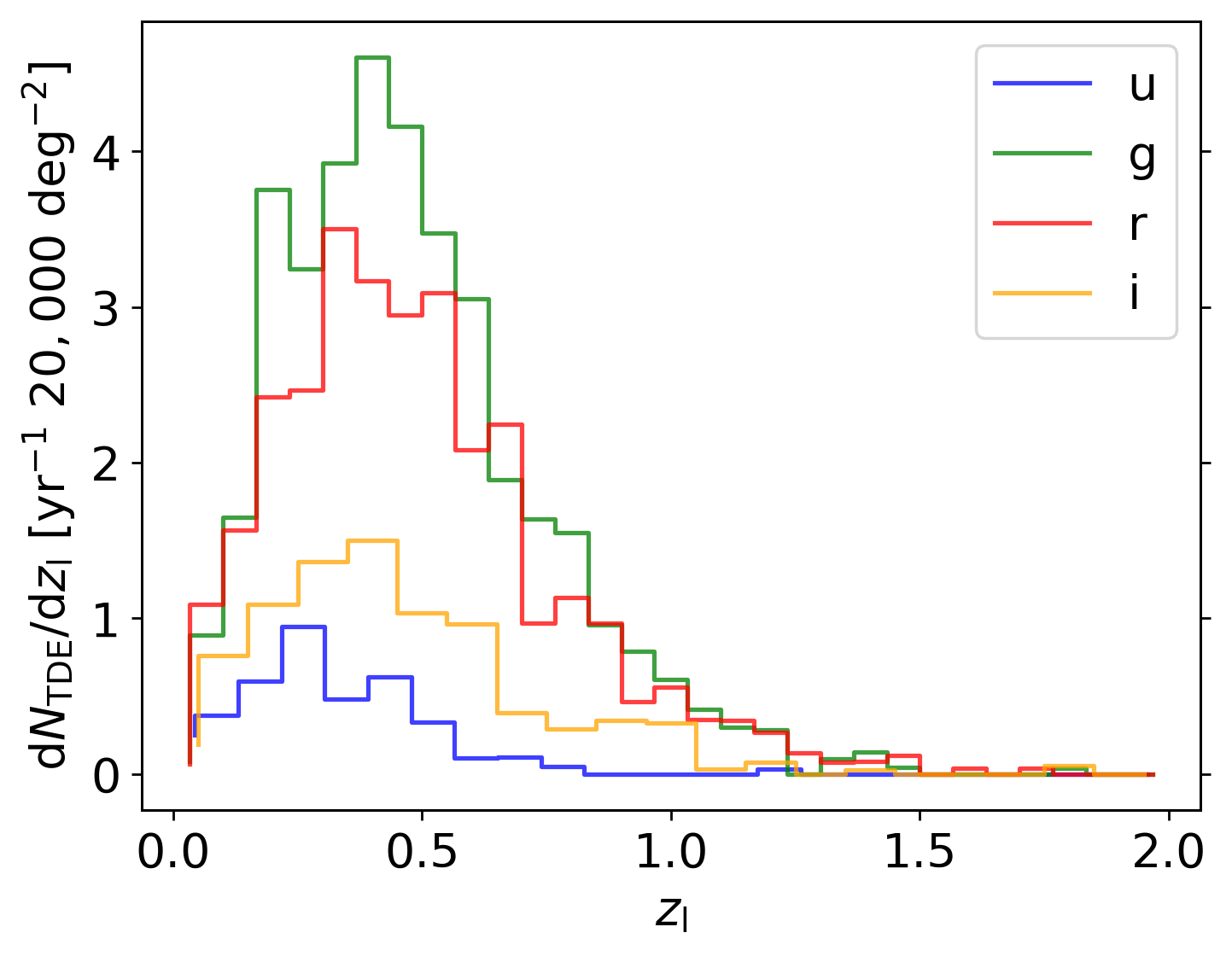}\label{fig*:zl_LSST}}\hfil
{\includegraphics[width=6cm]{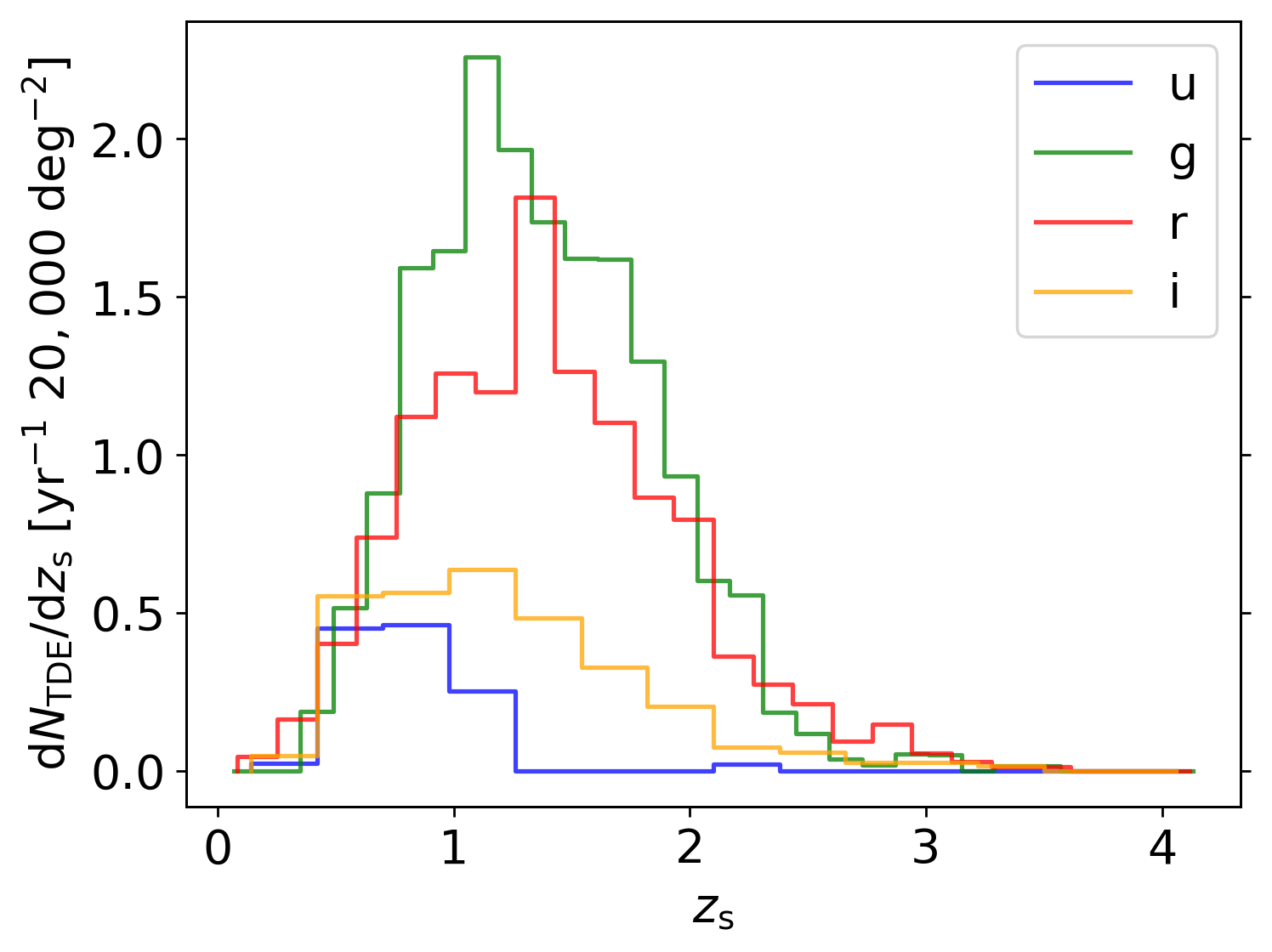}}\hfil 
{\includegraphics[width=6cm]{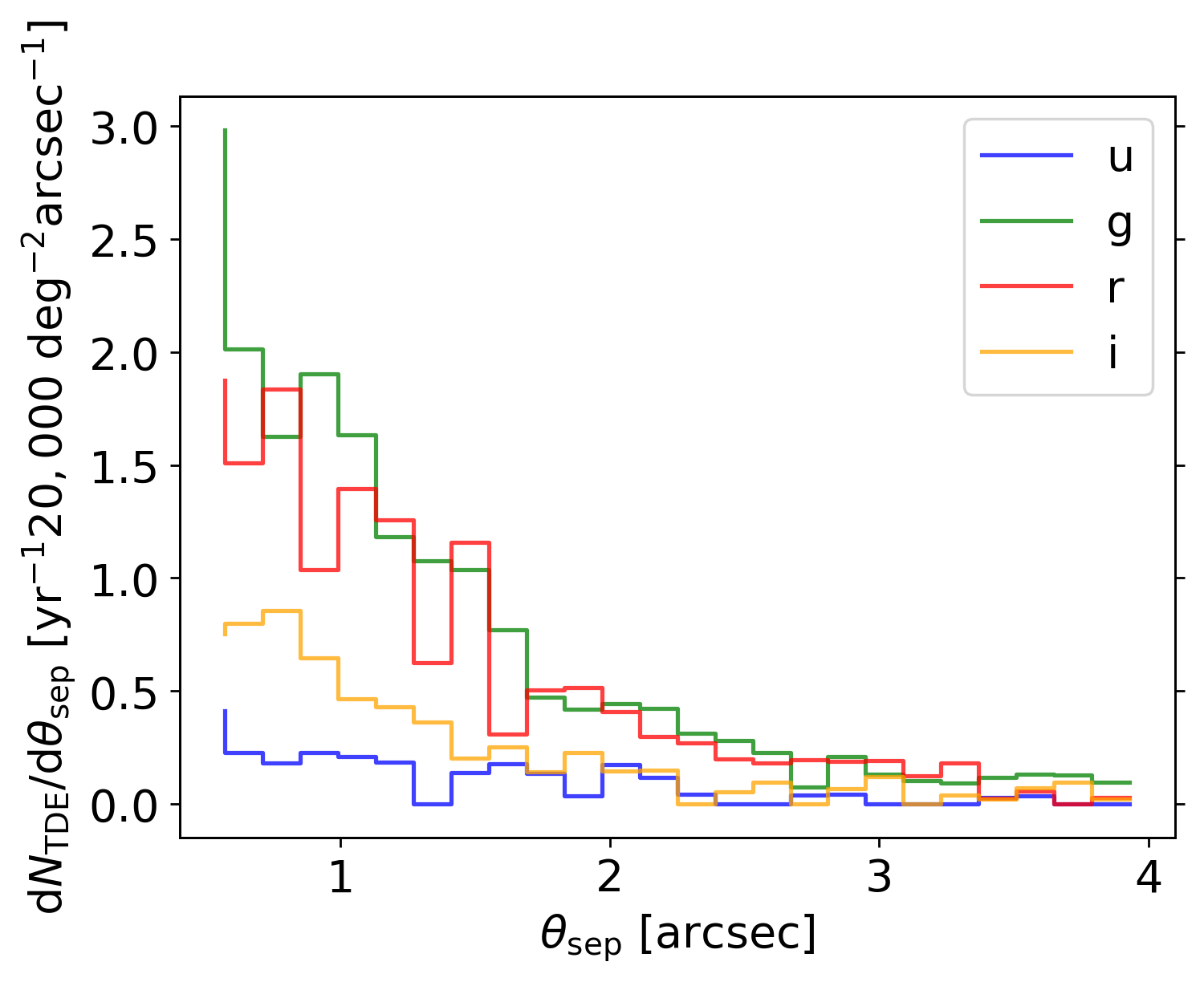}} 

{\includegraphics[width=6cm]{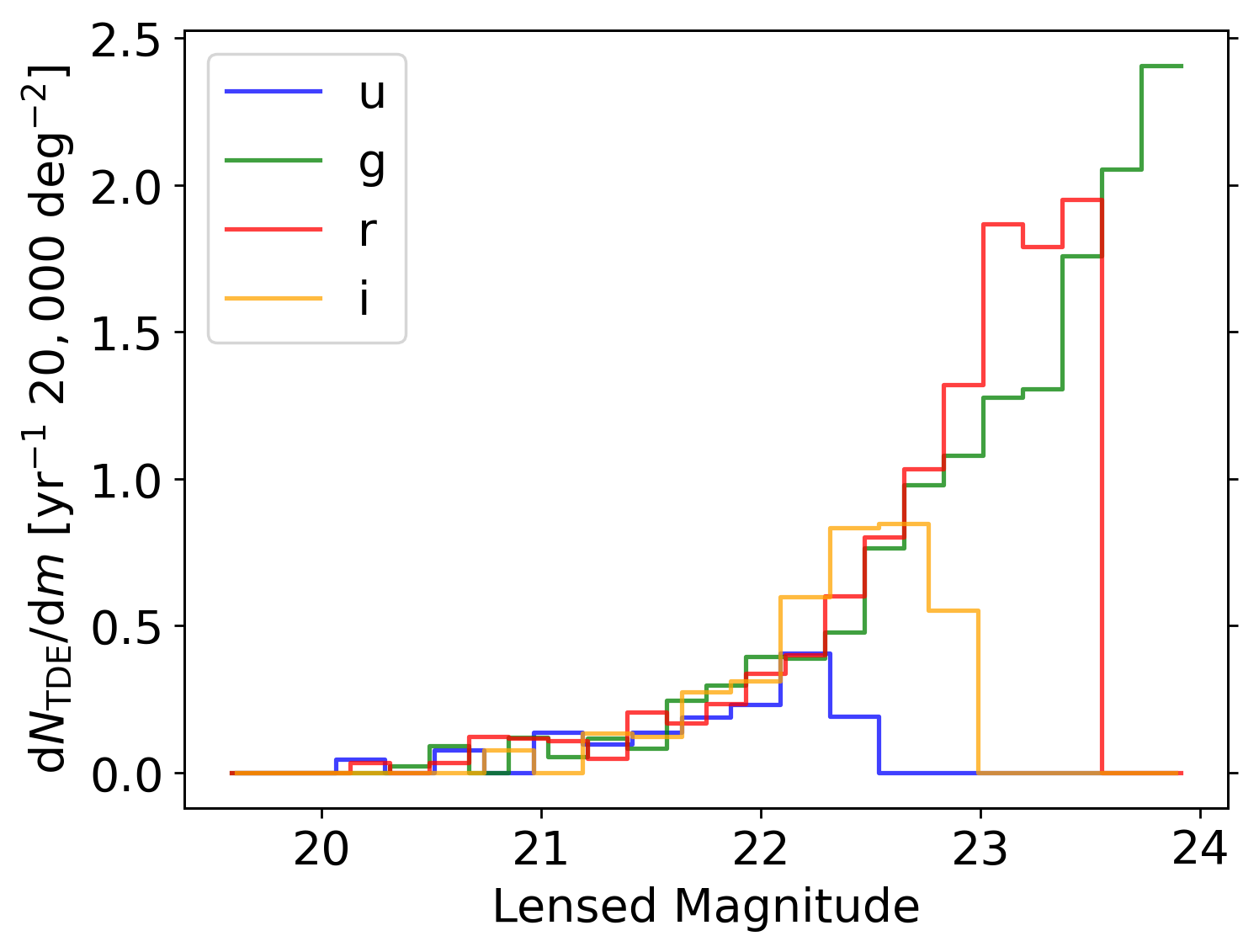}}\hfil   
{\includegraphics[width=6cm]{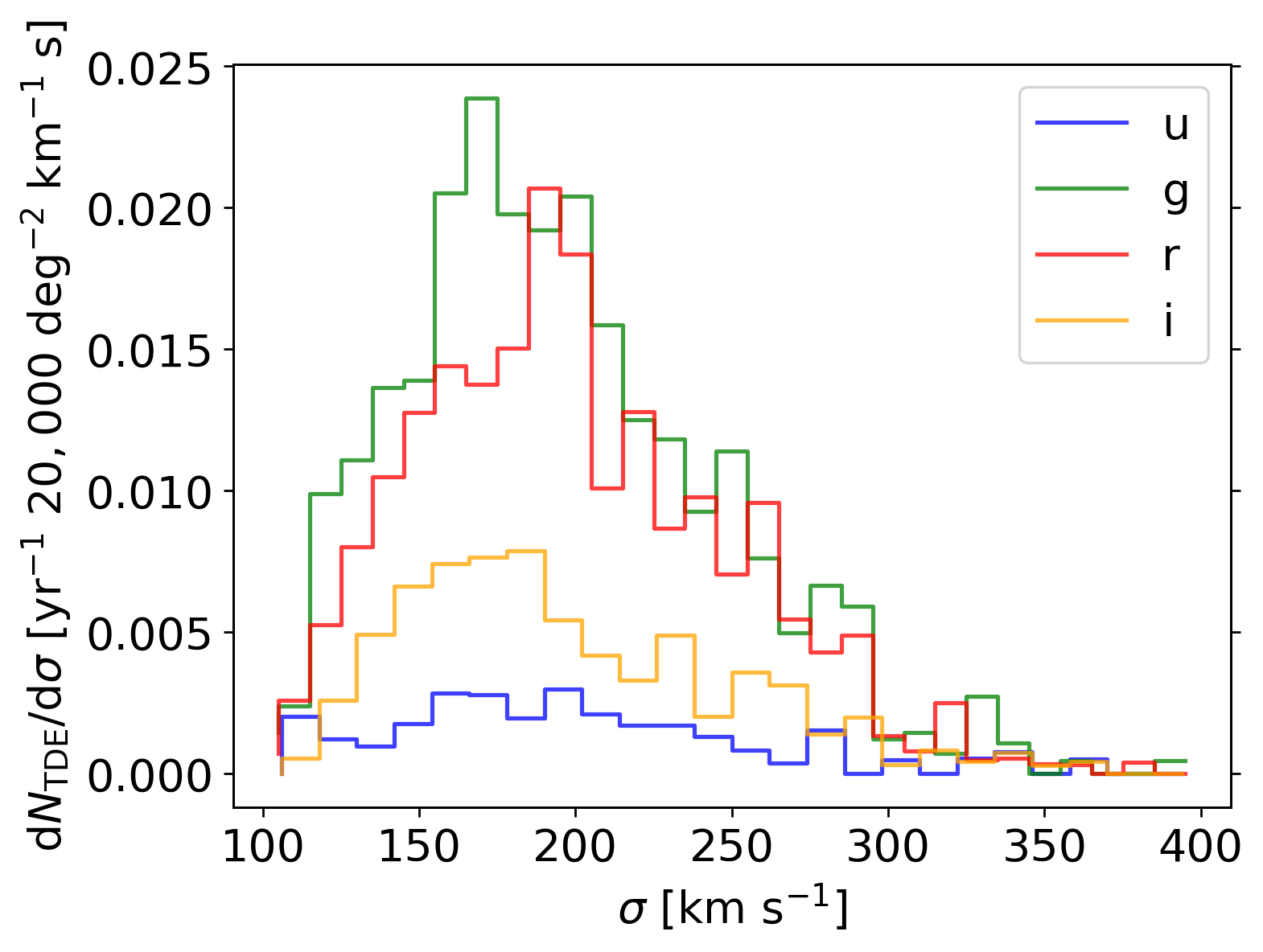}}\hfil
{\includegraphics[width=6cm]{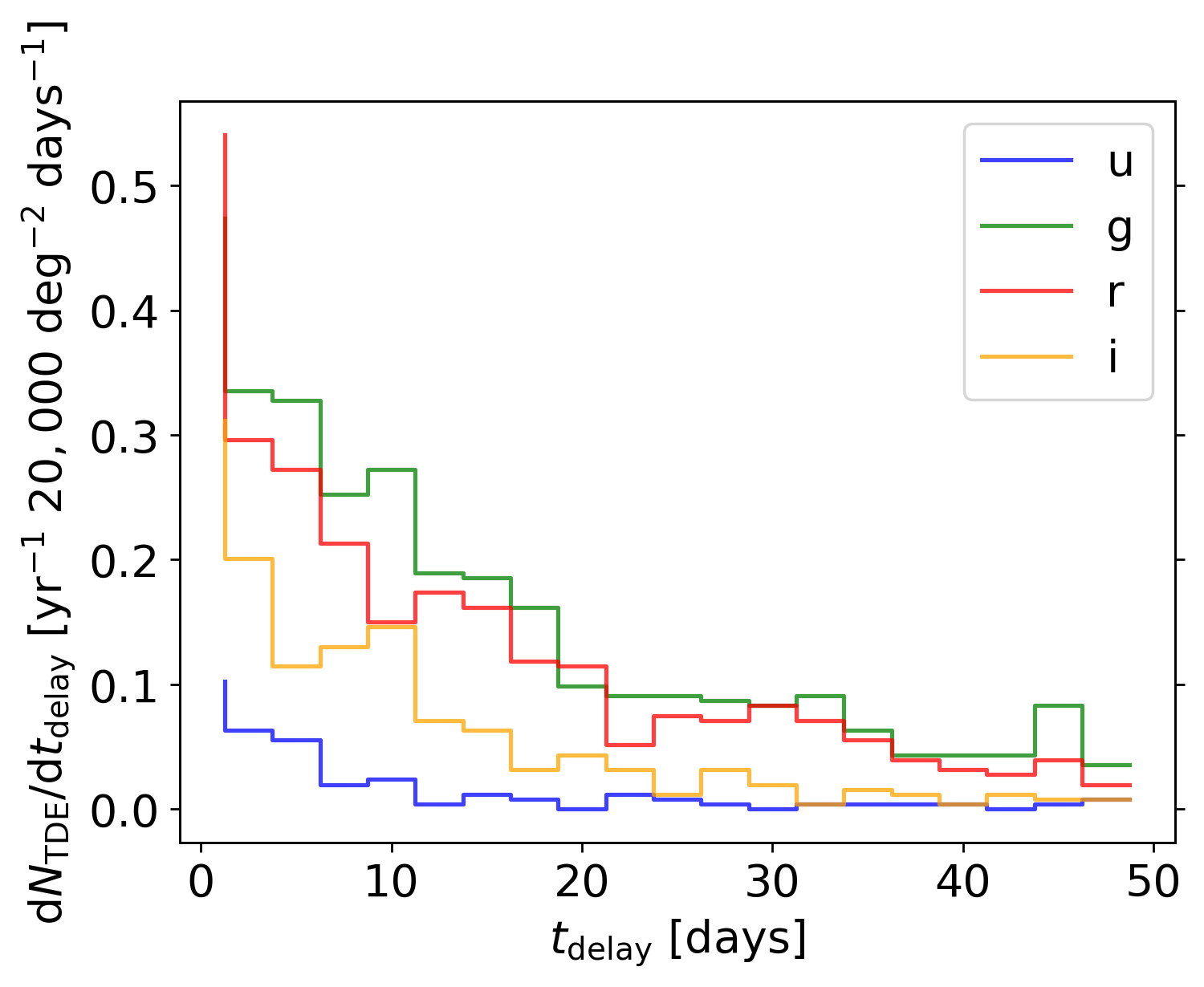}}
\caption{The distributions of the lensing properties of the strongly lensed $L_{2}$ systems from detections in each of the $u$, $g$, $r$ and $i$ bands after applying image separation cuts and LSST1 magnitude cuts. Top panels from left to right: lens galaxy redshift $z_{\rm l}$, TDE redshift $z_{\rm s}$, and image separation $\theta_{\rm sep}$.  Bottom panels from left to right: lensed magnitude (magnitude of the fainter image in a double system or the third brightest image in a quadruple system), stellar velocity dispersion $\sigma$ of the lens galaxy, and time delays $t_{\rm delay}$ of all images in the lensed systems.
}
\label{fig:L2_lens_properties}
\end{figure*}

In Figure \ref{fig:L2_lens_properties}, we show the properties of the $L_{2}$ lensing systems for $u$, $g$, $r$ and $i$ bands that would still be detected after applying LSST1 magnitude cuts. These panels indicate that $g$ and $r$ bands result in greater detections than the other bands, but we can also see that the properties of the systems detected by these two bands are very similar. The source redshift corresponding to the peak of the $u$ band distribution is at $z_\mathrm{s}=1.0$, but in the \textit{top middle} panel we observe the peak shifting to further redshifts, as in the unlensed case, such that the peaks fall at $z_\mathrm{s}=1.2$ and $z_\mathrm{s}=1.4$ for $g$ and $r$ bands, respectively. The image separation distributions are similar across all bands; the majority of strongly lensed systems have smaller image separations with $\theta_\mathrm{sep} \lesssim 1\arcsec$, and as the image separations increase further the distributions fall off as systems with larger separations are more unlikely. The lens velocity dispersion distributions are also similar across all four bands.

\begin{figure*}[ht]
\centering
{\includegraphics[width=6cm]{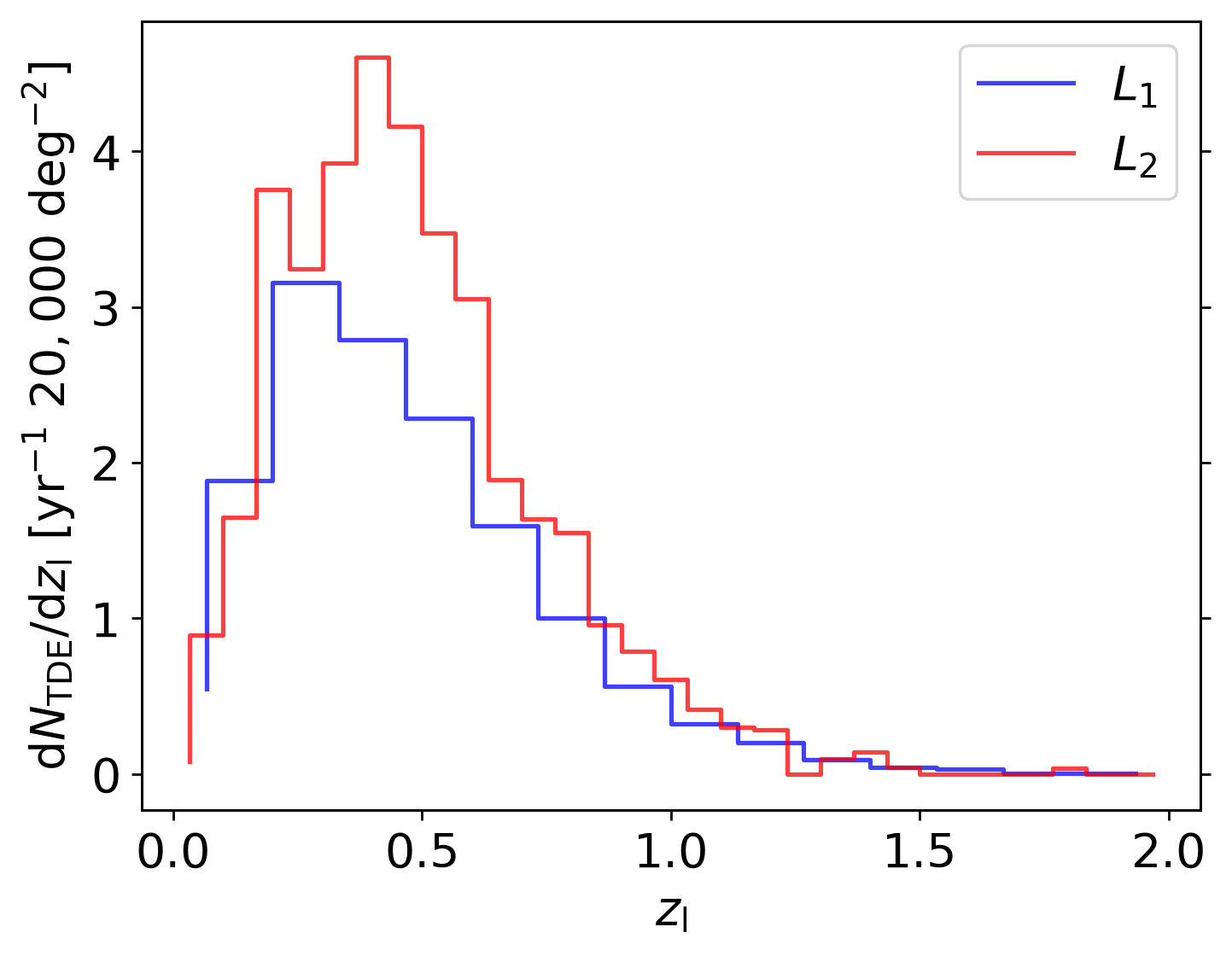}}\hfil
{\includegraphics[width=6cm]{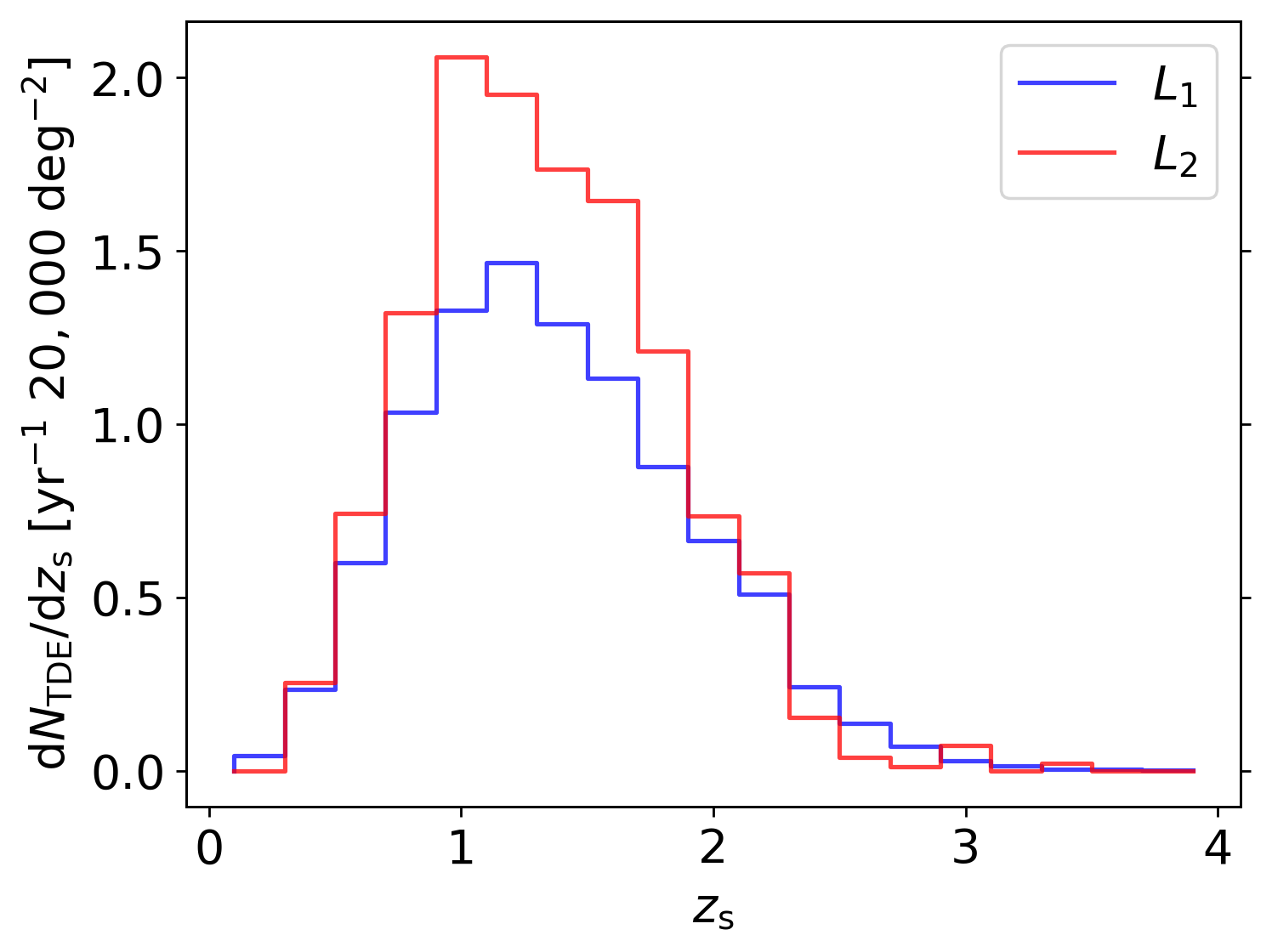}}\hfil 
{\includegraphics[width=6cm]{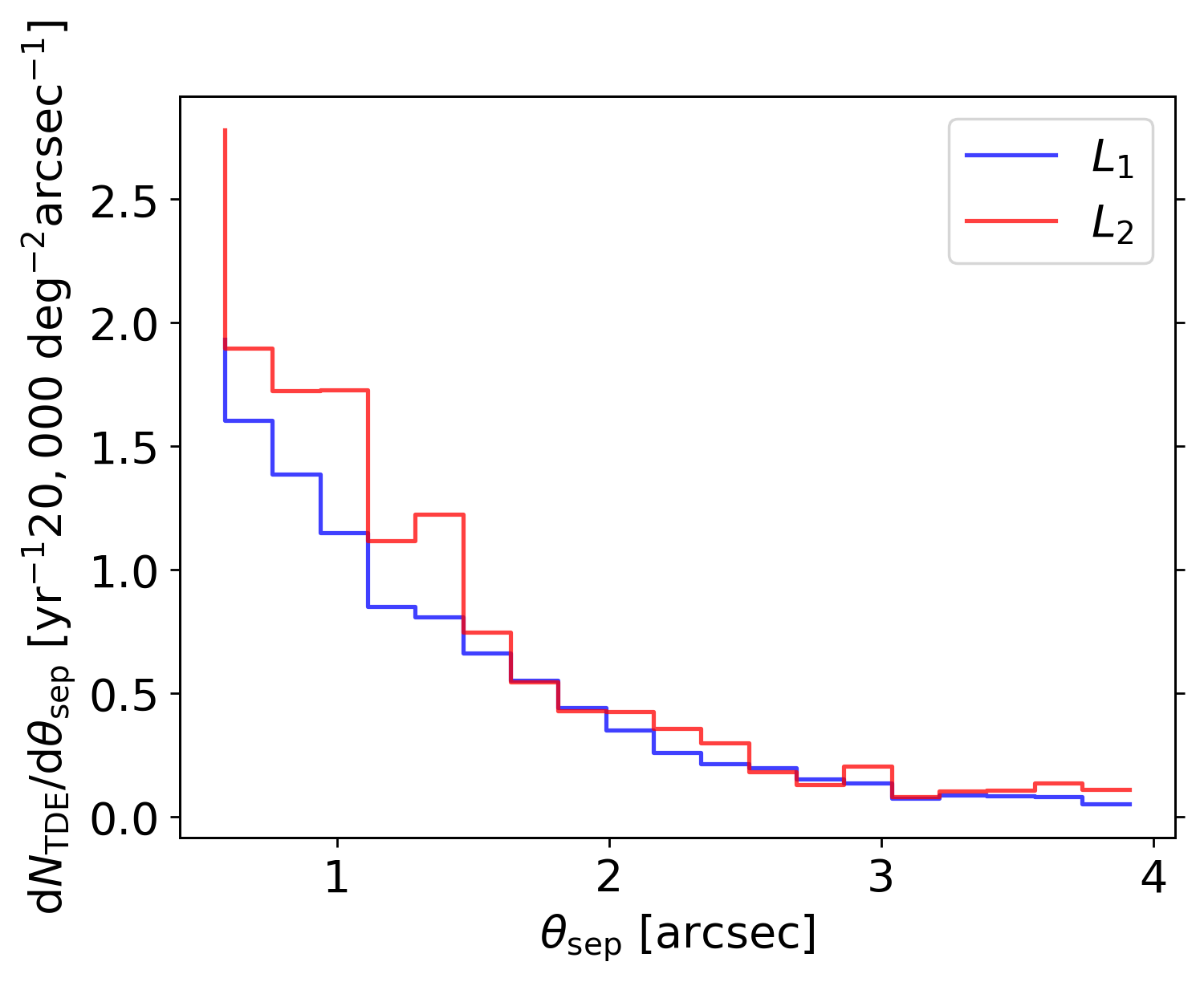}} 

{\includegraphics[width=6cm]{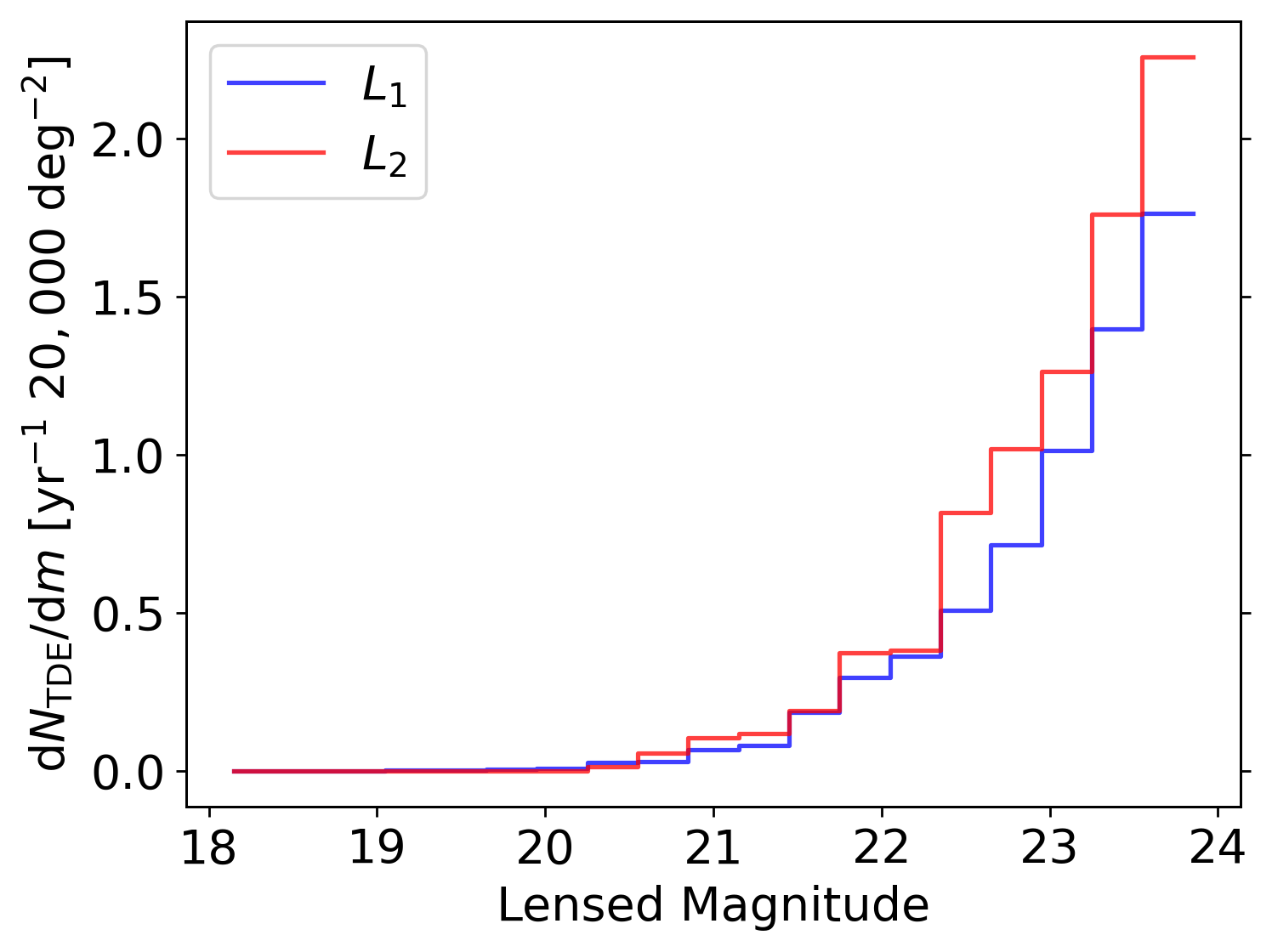}}\hfil   
{\includegraphics[width=6cm]{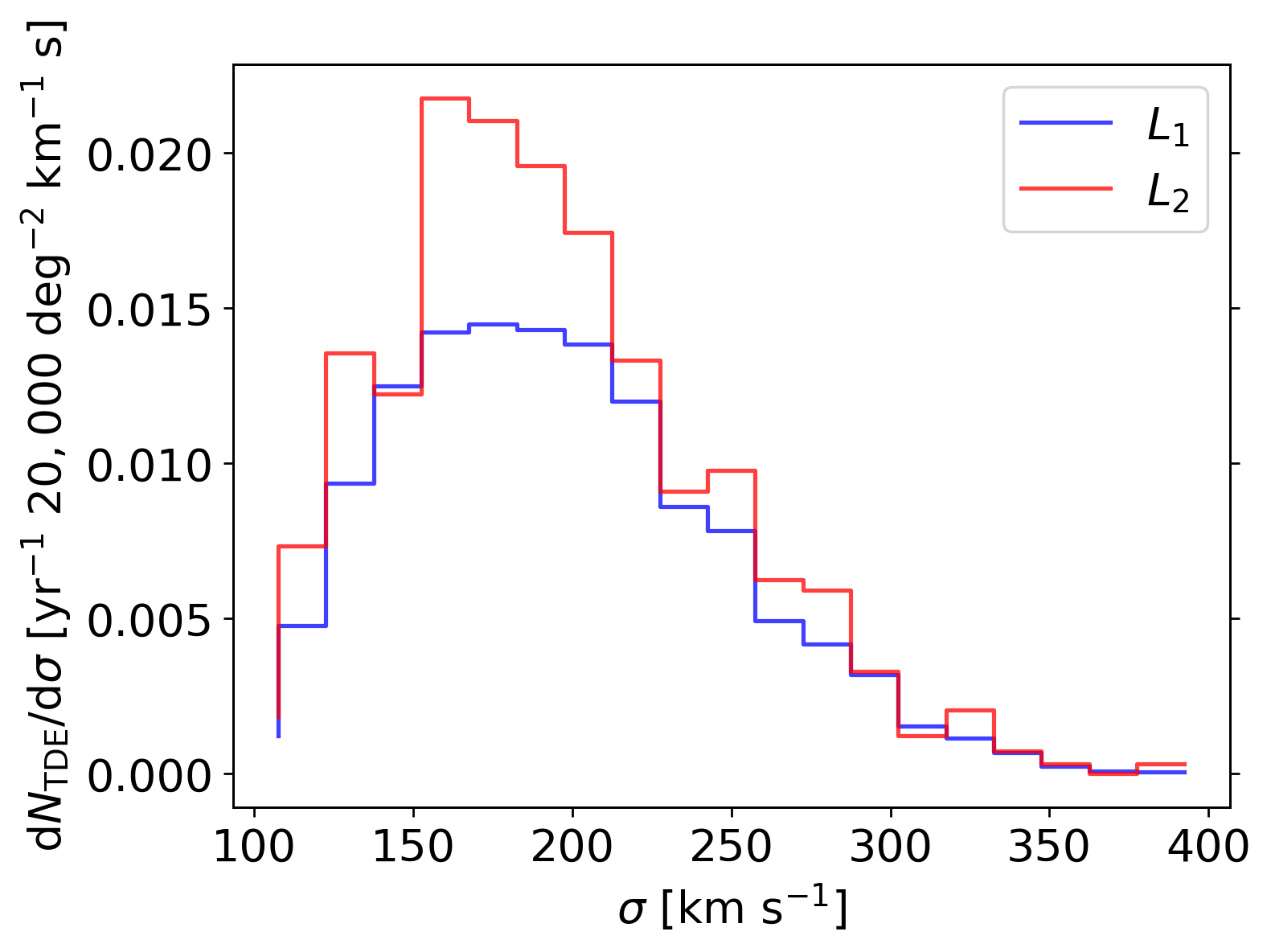}}\hfil
{\includegraphics[width=6cm]{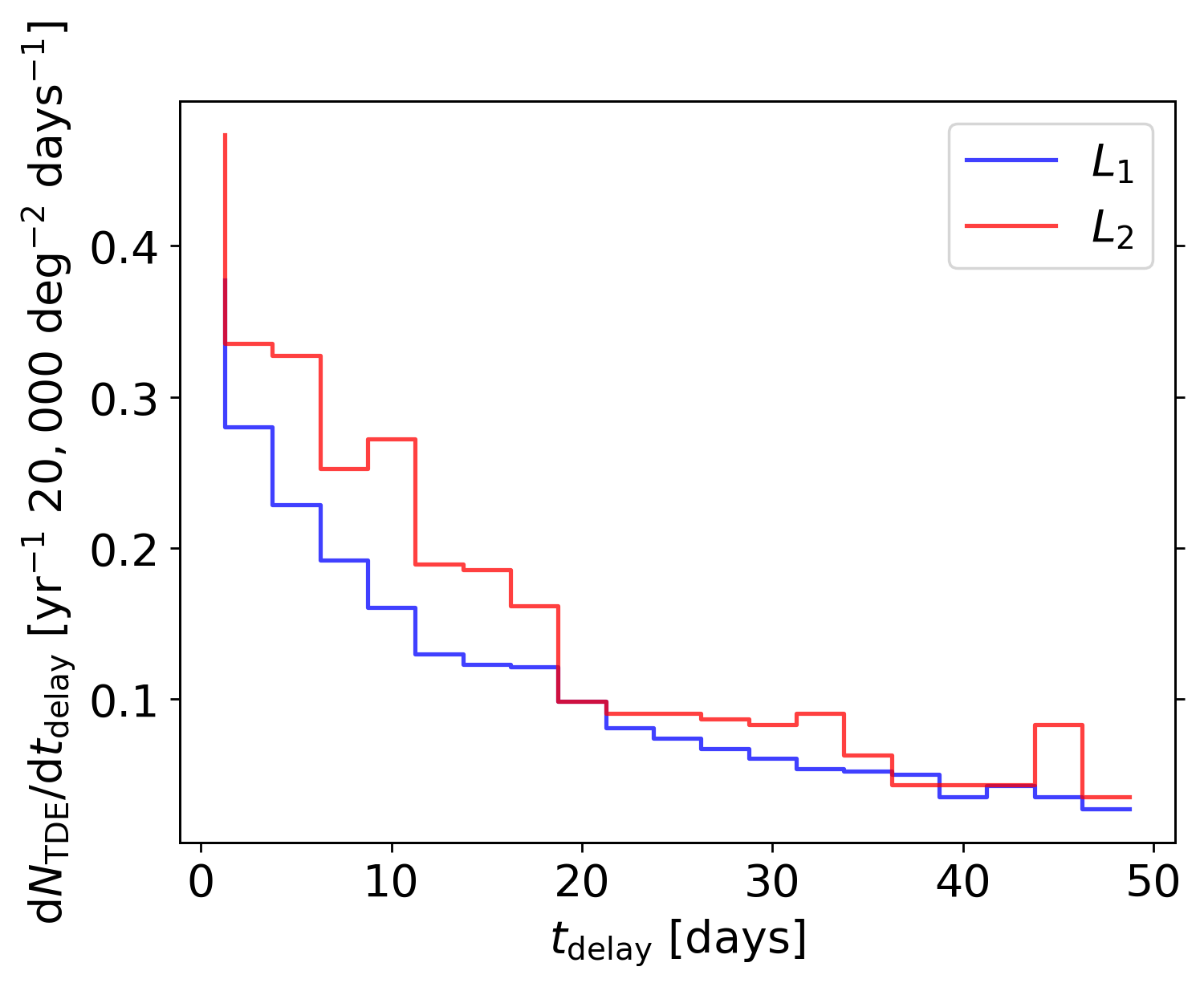}}
\caption{The distributions of the lensing properties of the strongly lensed $L_{1}$ and $L_{2}$ systems for $g$ band after applying image separation cuts. These distributions assume LSST1 magnitude cuts, and the panels are arranged in the same way as in Fig.~\ref{fig:L2_lens_properties}.
In general, the shapes of the distributions of the lensing properties are similar between $L_{1}$ and $L_{2}$.}
\label{fig:L1vsL2_lens_properties}
\end{figure*}

The $g$ band lensing properties for $L_{1}$ and $L_{2}$ are presented together in Figure \ref{fig:L1vsL2_lens_properties} to compare the results from these two luminosity models. We find that the distributions for these models are very comparable with the primary difference being that $L_{2}$ produces higher rates in general. 
We do not see a larger difference in the lensed magnitude panel, despite the $L_{1}$ luminosity spanning a larger magnitude range, because we have limited $L_{1}$ based on the Eddingtion luminosity.

\section{Discussions}
\label{sec: discussion}
In this work, we employ a BH mass function and a TDE rate per galaxy that are valid for the local universe, and we assume that they do not evolve with redshift. As seen in the lensed source redshift distribution in Section \ref{sec:lens properties}, the peak of the $L_{2}$ source redshift distribution for $u$ band is at $z_\mathrm{s}=1.0$, and all of the other bands peak at an even higher redshift. Thus, it would be useful to consider other BH mass functions and TDE occurrence rates that would be more relevant at these redshifts as these could affect the rates, which we will investigate in the future.

Additionally, we assumed a constant temperature for the TDEs. While this is observationally motivated, there is a theoretical temperature dependence for $L_{2}$ that was not included in this study. This would be an important path for future work to assess how this temperature dependence influences both the unlensed and strongly lensed detection rates. 

In general, we observe that, given the LSST1 magnitude cutoff, the $L_{2}$ $i$ and $u$ strongly lensed detections peak at a source redshift comparable to that of lensed supernovae which peak at $z_\mathrm{s} \lesssim 1$ \citep{Oguri_2010,Wojtak_2019} while the $g$ and $r$ bands peak at slightly higher redshifts, and this also applies for $L_{1}$. Thus, given the detection band, strongly lensed TDEs are expected to be detected at redshifts ${\sim}0.5$ higher than that of lensed supernovae.
As shown in both Figures \ref{fig:L2_lens_properties} and \ref{fig:L1vsL2_lens_properties}, the image separation $\theta_\mathrm{sep}$ is peaking toward the lower values for $L_{1}$ and $L_{2}$. If we were to retain lensed detections with $\theta_\mathrm{sep} < 0.5\arcsec$, we would subsequently obtain more strongly lensed systems. However, many of these systems would likely be unresolved by ground-based imaging surveys such as LSST. Nonetheless, there are methods to find such unresolved systems \citep[e.g.,][]{1996MNRAS.282..530G,2022ApJ...927..191B}, so it is useful to note the abundance of lensed systems within this region of $\theta_\mathrm{sep}$. We provide these full image separation distributions including detections with $\theta_\mathrm{sep} < 0.5 \arcsec$ in Appendix \ref{sec:imsep_distributions}.

Recently, \citet{Chen+2024} estimated the detection rates of strongly lensed TDEs by computing the probability of strong lensing (also known as the ``lensing optical depth'') of TDE light curves. They assumed that TDEs are powered by super-Eddington winds generated from an accretion disk that forms after the debris promptly circularizes \citep{Strubbe+2009, LodatoRossi2011}, which differs from our luminosity models. Despite differences in the assumptions for the luminosity function and the methodology of lensing calculations, our estimates for the lensed TDE detection rates by LSST are quite similar to theirs in the case where stars are tidally and fully destroyed at the largest possible distances from the black hole. However, their estimates are very sensitive to the assumption for the distance at which stars are destroyed, directly affecting the luminosity of TDEs, whereas our estimates are robust against assumptions for the luminosity. In addition, \citet{Chen+2024} considered all lensed TDEs with lensed image separations $\theta_\mathrm{sep}$ above $0.1\arcsec$, which provide higher lensed TDE rates compared to our $\theta_\mathrm{sep} > 0.5\arcsec$ limit.  In Appendix \ref{sec:imsep_distributions}, we show also our lensed TDE rates and properties where we drop the $0.5\arcsec$ criterion and consider all possible image separations.  The resulting rates of lensed TDEs per year from their and our independent analyses further corroborates the bright prospect of detecting lensed TDEs in the era of LSST.

\section{Summary}
\label{sec: summary}
In this paper, we present the estimated unlensed TDE detection rates for three observational survey magnitude cutoffs for two TDE luminosity models, $L_{1}$ and $L_{2}$, at five distinct temperatures. We show how the unlensed detection rates change for both luminosities with $m_\mathrm{lim,peak}$ and temperature. We also estimate the strongly lensed TDE detection rates for one of the LSST detection thresholds at the chosen temperature that best corresponds with current TDE observations, and we similarly show how these lensed detection rates change with $m_\mathrm{lim,peak}$. 

We find that $L_{1}$ and $L_{2}$ produce comparable unlensed detection rates. As $m_\mathrm{lim,peak}$ increases, the detection rates similarly increase as this means that the survey is probing at further depths and is then capable of detecting more TDEs. At lower magnitude cutoffs, the $u$ and $g$ bands result in the higher detections, but, as the magnitude cutoff increases, the TDEs are at further redshifts in general which enables the $i$ and $r$ bands to produce more detections which is well shown in Figure \ref{fig:ntde_allM}. We observe how the LSST1 and ZTF survey detections change with temperature; we find that the unlensed $L_{1}$ and $L_{2}$ LSST1 detections peak at $T=2 \times 10^{4}$ K before decreasing, and the ZTF detections all decrease with temperature. We find that, for both $L_{1}$ and $L_{2}$, low-mass black holes ($\mbh\lesssim 10^{6}$M$_{\odot}$) contribute the most to the rates, resulting in an overestimate of the ZTF detection rates by about an order of magnitude.

A close investigation for the overlap of unlensed detections across the four bands shows that the $g$ band will yield the highest number of TDEs at $T \geq 2 \times 10^{4}$ K, while $r$ band will yield the highest number at $T = 1 \times 10^{4}$ K. However, for $L_1$ with $T = 1 \times 10^{4}$ K, the $g$ band may still miss some detections from some of the other remaining bands. To maximize the number of TDE detections, we advocate for using the combination of  $g$, $r$ and $i$ bands in LSST-like surveys, with $g$ and $r$ bands being the primary bands; if observing resources are limited, then we advocate for prioritizing the four bands in the following order: $g$, $r$, $i$ and then $u$.

Similar to the unlensed TDEs, we find that $L_{1}$ and $L_{2}$ result in very comparable strongly lensed detection rates with $L_{2}$ consistently producing higher rates. Assuming the LSST1 cutoff for the peak TDE magnitude given a TDE temperature of $T=2 \times 10^{4}$ K for the entire duration of LSST (4.5 effective years), the estimated strongly lensed detection rates for $L_{1}$ are ($u$, $g$, $r$, $i$) = (1.1, 8.7, 7.1, 3.2), and these detection rates for $L_{2}$ are ($u$, $g$, $r$, $i$) = (1.5, 11, 9.1, 3.8). These rates indicate that we may detect $\sim 1$ lensed event for every $10^{4}$ unlensed detections, which is robust against the assumptions for the event luminosity. 
For $L_{2}$ with the LSST1 magnitude cutoff, $g$ and $r$ are expected to observe lensed TDEs at higher redshifts than $i$ and $u$.

Our study shows that strongly lensed TDEs will likely be discovered in the coming years as wide-field imaging surveys such as LSST start operating. This will open a new window to study TDEs via the strong lensing and microlensing effects.

\FloatBarrier

\begin{acknowledgements}

We thank S. de Mink for the useful discussions. KS, SHS and SH thank the Max Planck Society for support through the Max Planck Research Group and Max Planck Fellowship for SHS. This project was supported through a Fulbright grant of the German-American Fulbright Commission. This project has received funding from the European Research Council (ERC) under the European Union's Horizon 2020 research and innovation
programme (LENSNOVA: grant agreement No 771776). LD acknolwedges the support from the National Natural Science Foundation of China (HKU12122309) and the Hong Kong Research Grants Council (HKU17304821, HKU17314822, HKU27305119). This work was supported by JSPS KAKENHI Grant Numbers JP22H01260, JP22K21349, JP19KK0076.

\end{acknowledgements}

\bibliographystyle{aa}
%\bibliography{TDE_bib}

\FloatBarrier

\clearpage
\appendix
\section{Full expression for $\psi_{\rm TDE}(m,z)$}
\label{sec:lf_expression}
The TDE luminosity function in terms of magnitude can be expressed by, 
\begin{align}
&\psi_{\rm TDE}\left(m,z\right)=\psi_{\rm TDE}
\left(L_{i}\left(m\right)\right)
\frac{\mathrm{d}L_{i}\left(m\right)}{\mathrm{d}m},\nonumber\\
    &=10^{-13.22}  \left(\frac{M_{\rm BH}\left(L_{i}\left(m\right)\right)}{10^{6}\,\Ms}\right)^{-1.24}10^{ -\left[\frac{M_{\rm BH}\left(L_{i}\left(m\right)\right)}{128\times 10^{7}\,\Ms}\right]^{\left(\frac{1}{\ln(10)}\right)}}\nonumber\\
    &\times \frac{\mathrm{d}M_\mathrm{BH}\left(L_{i}\left(m\right)\right)}{\mathrm{d}L_{i}\left(m\right)},
\end{align}
%\sh{should be labeled as one Equation A.1}\taeho{fixed}
with 
%$L_{i}=1,2$, 
$i=1,2$,
based on the luminosity model applied, defined as,
\begin{align}
    L_{i}\left(m\right) &=\Xi(T, \nu_{\rm max}, \nu_{\rm min},z)^{-1} 10^{\frac{m + 48.6}{-2.5}}.
\end{align}
The black hole mass $M_{\rm BH}\left(L_{i}\left(m\right)\right)$ for $L_{1}$ and $L_{2}$ can be obtained from Equations (\ref{eq:L1_simplified}) and (\ref{eq:L2}), respectively. These expressions can be given explicitly by,
\begin{align}
M_{\rm BH}\left(L_{1}\left(m\right)\right) &=8.43 \times 10^{6}\Ms\left(\frac{L_{1}\left(m\right)}{10^{44}~{\rm erg~ s}^{-1}}\right)^{-2},
\end{align}
\begin{align}
M_{\rm BH}\left(L_{2}\left(m\right)\right) &=8.26 \times 10^{5}\Ms\left(\frac{L_{2}\left(m\right)}{10^{44}~{\rm erg~ s}^{-1}}\right)^{-6}.
\end{align}

\section{Unlensed Magnitude Relations}
\label{sec:appendix_mag_conversion}
%\sherry{How about using ``Magnitude relations" rather than ``Magnitude conversions" throughout?}

\begin{figure}[h!]
    \centering
\includegraphics[width =8.cm]{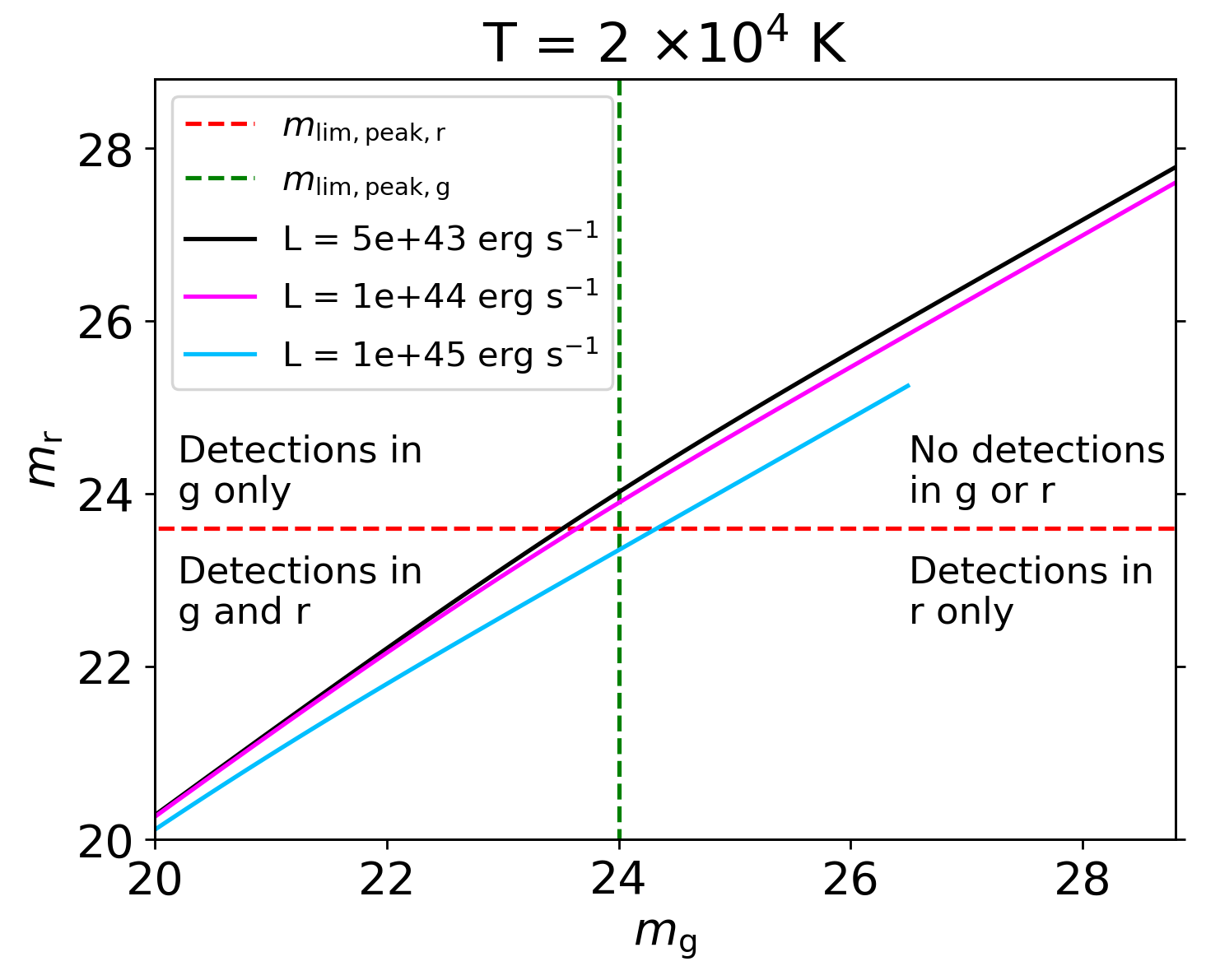}
\includegraphics[width =8.cm]{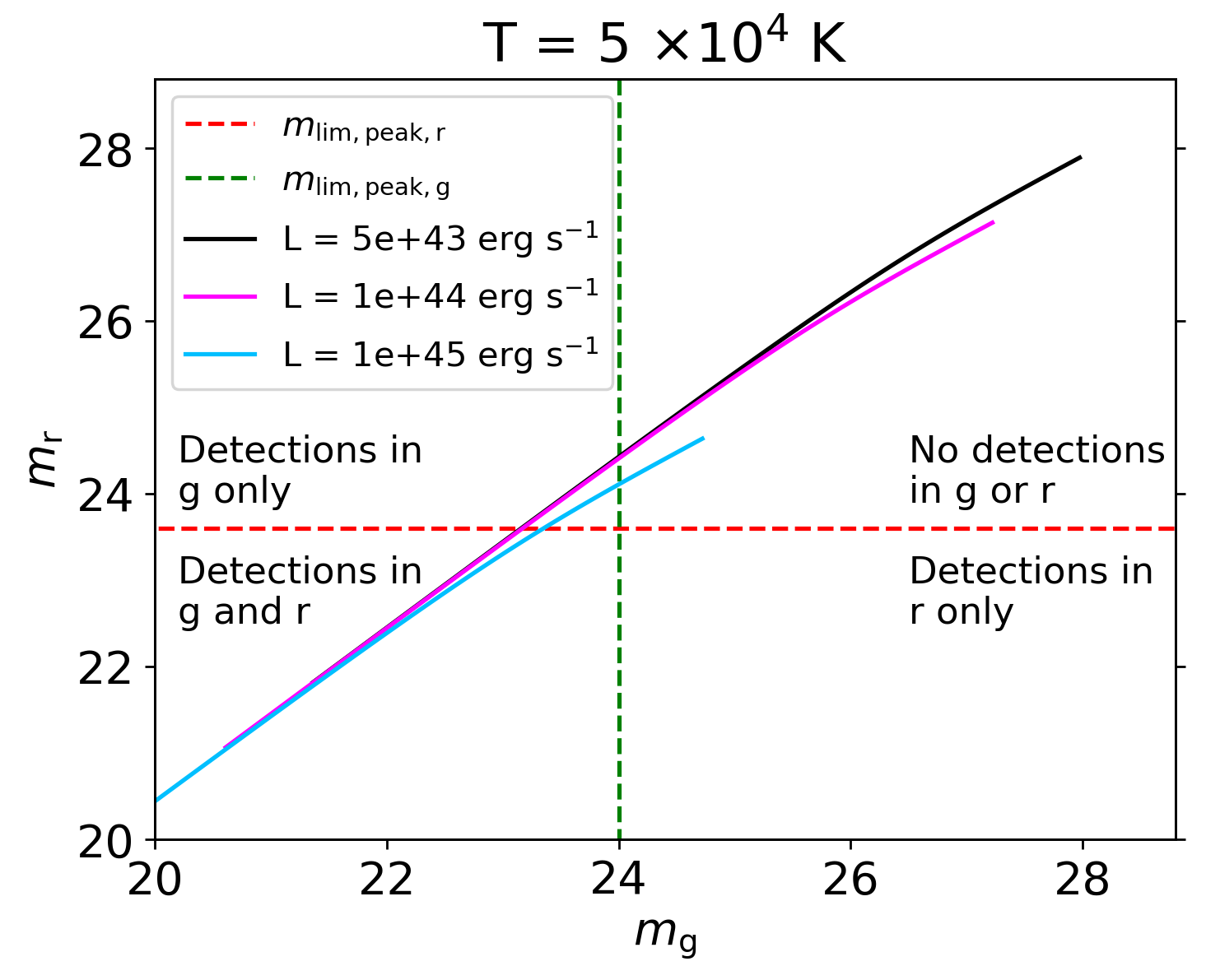}
\caption{The unlensed magnitude relationship between $r$ and $g$ bands at $T=2$ and $5 \times 10^{4}$ K. Assuming a redshift range of $z = 0.1 - 5$, three luminosity cases of $L=5 \times 10^{43}$, $1 \times 10^{44}$ and $1 \times 10^{45} ~\mathrm{erg}~\mathrm{s^{-1}}$ are shown to represent the luminosity span across $L_{1}$ and $L_{2}$, and the $m_\mathrm{lim,peak}$ values for $r$ and $g$ bands are given by the red and green dashed lines, respectively.
}
\label{fig:mag_conversion}
\end{figure}

Figure \ref{fig:mag_conversion} shows the relation between the $r$ band magnitude and the $g$ band magnitude for two example cases, one with partial overlappings in detections between bands (\textit{top}, $T=2 \times 10^{4}$ K) and one with a complete hierarchy in band detections (\textit{bottom}, $T=5 \times 10^{4}$ K). We take the three constant luminosity values, i.e., $L=5\times10^{43}~{\rm erg~s}^{-1}$, $10^{44}~{\rm erg~s}^{-1}$, and $10^{45}~{\rm erg~s}^{-1}$, and vary the redshift to examine the magnitude conversion. Since the relation between the two different bands solely depends on the relation between $L$ and $m$,  for given $L$, the shape of the line is independent of the luminosity models (i.e., $L_{1}$ and $L_{2}$). However, each luminosity model covers a different range of $L$: the two smaller luminosities are covered by the luminosity range for both $L_{1}$ and $L_{2}$ while the largest luminosity ($10^{45}~{\rm erg~s}^{-1})$ is only relevant for $L_{1}$. For simplicity in this appendix, we do not limit $L_1$ to the Eddington luminosity, as this does not affect the explanation below of the hierarchy different bands for TDE detection.

In the \textit{top} panel of \ref{fig:mag_conversion}, all TDEs within the luminosity range of $L=5 \times 10^{43} - 1 \times 10^{44} ~\mathrm{erg}~\mathrm{s^{-1}}$ that are brighter than the $g$ band magnitude are above the $r$ band magnitude (i.e., third quadrant, left side of the green vertical dashed line below the red horizontal dashed line), meaning these TDEs detected in $g$ band would be detected in $r$ band. On the other hand, there are events with $L=1 \times 10^{45} ~\mathrm{erg}~\mathrm{s^{-1}}$ whose magnitudes are still brighter than the $m_\mathrm{lim,peak,r}$, but fainter than the $m_\mathrm{lim,peak,g}$ (fourth quadrant), implying some detections found in $r$ are missed in $g$. So a full hierarchy between the two bands ($g\supset r$) has been established for the range of luminosity relevant for $L_{2}$. However, $L=1 \times 10^{45} ~\mathrm{erg}~\mathrm{s^{-1}}$, which is only relevant for $L_{1}$, reveals the opposite hierarchy ($r\supset g$). Hence, the inhomogeneous hiearchical relation accounts for the partial overlap observed for $L_{1}$ in Figure \ref{fig:l1_unlensed_hierarchy} and not seen in the $L_{2}$ hierarchy. Conversely, in the $\textit{bottom}$ panel, all TDEs within the luminosity range being assessed that are brighter than $m_\mathrm{lim,peak,r}$ will also be detected in $g$. For this range of luminosity, there exists a full hierarchy between $g$ and $r$ band detections ($g\supset r$). We observe this trend represented in Figures \ref{fig:l1_unlensed_hierarchy} and \ref{fig:l2_unlensed_hierarchy} by the green box for $g$ being taller than the red $r$ box. 

The annual unlensed detections for $L_{1}$ and
$L_{2}$ for each band at all temperatures are provided in Tables \ref{tab:converted_mag_l1} and \ref{tab:converted_mag_l2}. The magnitude of each TDE in the other three bands were computed to determine which bands the TDE would be detected in, and the detection counts in the other bands are listed under the `Detected Band' region of the tables.

\begin{table*}
\centering
\begin{tabular}{  c c c c c c} 
\hline
 & & & Detected Band\\
Temperature & Catalog Band & $u$ & $g$ & $r$ & $i$\\
\hline
\hline
1 $\times 10^{4}$K & u & 3,976* & 3,976 & 3,976 & 3,976 \\
1 $\times 10^{4}$K & g & 3,951 & 13,547* & 13,537 & 12,560 \\
1 $\times 10^{4}$K & r & 3,963 & 13,403 & 15,353* &13,228 \\
1 $\times 10^{4}$K & i & 3,960 & 12,645 & 13,413 &13,413* \\
\hline
2 $\times 10^{4}$K & u & 5,800* & 5,800 & 5,800 &5,701 \\
2 $\times 10^{4}$K & g & 5,717 & 17,892* & 13,971 & 7,247 \\
2 $\times 10^{4}$K & r & 5,703 & 13,962 & 13,962* & 7,208 \\
2 $\times 10^{4}$K & i & 5,594 & 7,164 & 7,164 &7,164* \\
\hline
3 $\times 10^{4}$K & u & 3,718* & 3,718 & 3,718 & 2,285 \\
3 $\times 10^{4}$K & g & 3,601 & 12,816*& 6,688 & 2,230 \\
3 $\times 10^{4}$K & r & 3,609 & 6,831 & 6,831* & 2,268 \\
3 $\times 10^{4}$K & i & 2,232 & 2,232 & 2,232 &2,232* \\
\hline
4 $\times 10^{4}$K & u & 1,830* & 1,830 & 1,830 & 736 \\
4 $\times 10^{4}$K & g & 1,904 & 7,122* & 2,782 & 734 \\
4 $\times 10^{4}$K & r & 1,815 & 2,713 & 2,713* & 710 \\
4 $\times 10^{4}$K & i & 787 & 787 & 787 &787* \\
\hline
5 $\times 10^{4}$K & u & 843* & 843 & 843 & 246 \\
5 $\times 10^{4}$K & g & 868 & 3,601* & 1,118 & 281 \\
5 $\times 10^{4}$K & r & 871 & 1,163 & 1,163* & 270 \\
5 $\times 10^{4}$K & i & 280 & 280 & 280 &280* \\
\hline
\end{tabular}
\caption{Annual unlensed $L_{1}$ detection rates per band that would be detected in the other three bands for all five temperatures given LSST1 magnitude cuts. The * symbols indicate the rate for the catalog bands.}
\label{tab:converted_mag_l1}
\end{table*}

\begin{table*}
\centering
\begin{tabular}{  c c c c c c} 
\hline
 & & & Detected Band\\
Temperature & Catalog Band & $u$ & $g$ & $r$ & $i$\\
\hline
\hline
1 $\times 10^{4}$K & u & 5,063* & 5,063 & 5,063 & 5,063 \\
1 $\times 10^{4}$K & g & 4,971 & 16,264* & 1,649 & 16,024 \\
1 $\times 10^{4}$K & r & 4,833 & 16,581 & 19,295* &17,211 \\
1 $\times 10^{4}$K & i & 4,862 & 16,028 & 16,770 &16,770* \\
\hline
2 $\times 10^{4}$K & u & 7,860* & 7,860 & 7,860 &7,853 \\
2 $\times 10^{4}$K & g & 7,591 & 23,199* & 18,463 & 9,392 \\
2 $\times 10^{4}$K & r & 7,505 & 18,785 & 18,785* & 9,355 \\
2 $\times 10^{4}$K & i & 7,412 & 9,195 & 9,195 &9,195* \\
\hline
3 $\times 10^{4}$K & u & 4,677* & 4,677 & 4,677 & 2,782 \\
3 $\times 10^{4}$K & g & 4,654 & 16,579*& 8,625 & 2,641 \\
3 $\times 10^{4}$K & r & 4,689 & 9,078 & 9,078* & 2,918 \\
3 $\times 10^{4}$K & i & 2,980 & 2,980 & 2,980 &2,980* \\
\hline
4 $\times 10^{4}$K & u & 2,380* & 2,380 & 2,380 & 880 \\
4 $\times 10^{4}$K & g & 2,411 & 9,071* & 3,645 & 920 \\
4 $\times 10^{4}$K & r & 2,232 & 3,133 & 3,133* & 843 \\
4 $\times 10^{4}$K & i & 932 & 932 & 932 &932* \\
\hline
5 $\times 10^{4}$K & u & 1,015* & 1,015 & 1,015 & 249 \\
5 $\times 10^{4}$K & g & 1,147 & 4,631* & 1,532 & 402 \\
5 $\times 10^{4}$K & r & 1,088 & 1,318 & 1,318* & 349 \\
5 $\times 10^{4}$K & i & 267 & 267 & 267 &267* \\
\hline
\end{tabular}
\caption{Same as Table~\ref{tab:converted_mag_l1}, but for $L_{2}$.}
\label{tab:converted_mag_l2}
\end{table*}

\section{Full $\theta_\mathrm{sep}$ Distributions}
\label{sec:imsep_distributions}

\begin{figure}[h!]
    \centering
\includegraphics[width =8.cm]{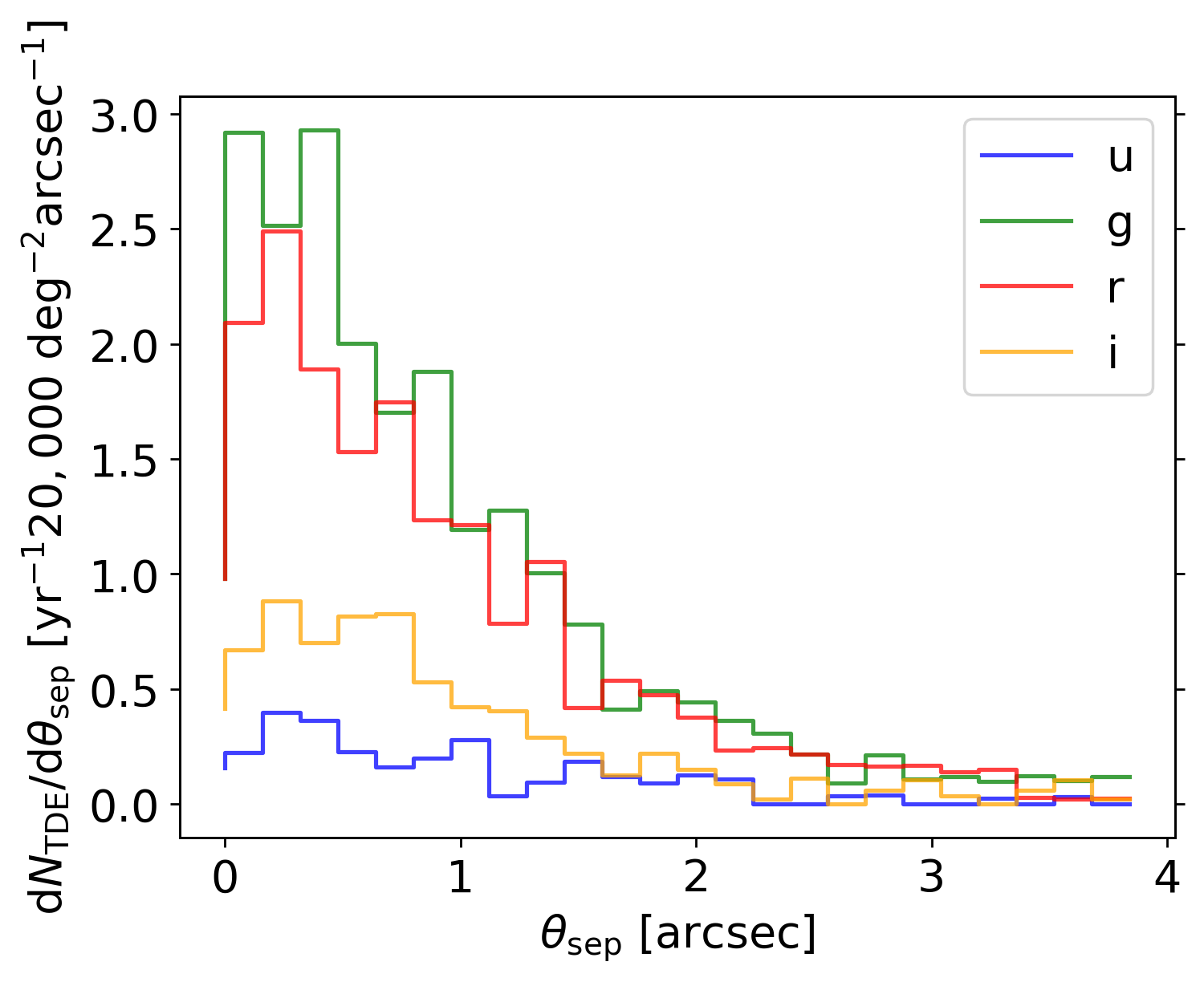}
\includegraphics[width =8.cm]{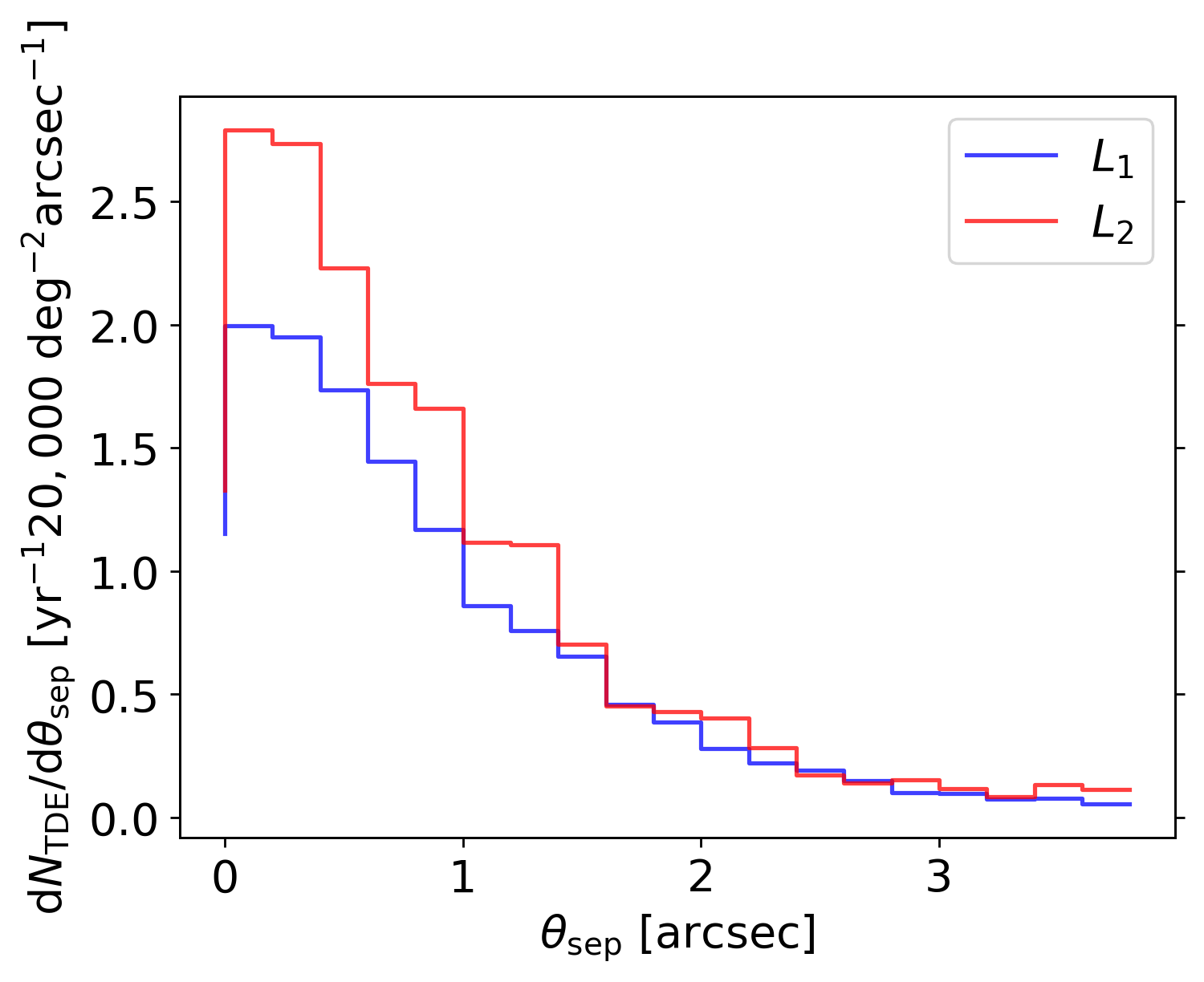}
\caption{The lensed TDE image separation distributions assuming LSST1 magnitude cutoffs without the condition that $\theta_\mathrm{sep} > 0.5 \arcsec$. The $\textit{top}$ panel shows the distributions for all four bands for $L_{2}$, and the $\textit{bottom}$ panel shows the $g$ band distributions for $L_{1}$ and $L_{2}$.
}
\label{fig:imsep_full}
\end{figure}

The full $\theta_\mathrm{sep}$ distributions for the lensed TDEs are shown in Figure \ref{fig:imsep_full}. These distributions do not require that $\theta_\mathrm{sep} > 0.5 \arcsec$, so these distributions provide an understanding on the number of systems that would likely be unresolvable by ground-based imaging surveys.

\end{document}